\documentclass[useAMS,usenatbib, usegraphicx]{mn2e}

\usepackage{graphicx}
\usepackage{hyperref}
\usepackage{amsmath,amssymb}
\usepackage{amsfonts}
\usepackage{xspace} 
\usepackage[usenames]{color}
\usepackage{dcolumn} 
\usepackage{bm} 
\usepackage{epsfig}
\usepackage{aas_macros}

\usepackage{ulem}
\definecolor {darkgreen}{rgb}{0.2,0.7,0.2}

\newcommand{\be}{\begin{equation}}
\newcommand{\ba}{\begin{eqnarray}}
\newcommand{\ee}{\end{equation}}
\newcommand{\ea}{\end{eqnarray}}

\newcommand{\SMBH}{\bullet}
\newcommand{\SCO}{\rm s}

\newcommand{\GW}{{\mbox{\tiny GW}}}

\newcommand{\ISCO}{{\mbox{\tiny ISCO}}}

\newcommand{\Msun}{\,{\rm M_\odot}}

\newcommand{\D}{\mathrm{d}}
\newcommand{\C}{c}

\newcommand{\R}{{\bar{r}}}

\newcommand{\G}{G}

\newcommand{\W}{\mathcal{W}}

\title[Gas pile-up, gap overflow, and Type 1.5 migration: SMBH binaries]{ Gas pile-up, gap overflow, and Type 1.5 migration in circumbinary
 disks:  application to supermassive black hole binaries}

\author[Kocsis, Haiman, \& Loeb]{Bence Kocsis$^{1,3}$\thanks{E-mail: bkocsis@cfa.harvard.edu},
Zolt\'an Haiman$^{2}$\thanks{E-mail: zoltan@astro.columbia.edu}
and Abraham Loeb$^{1}$\thanks{E-mail: aloeb@cfa.harvard.edu}
\\$^{1}$Harvard-Smithsonian Center for Astrophysics, 60 Garden St., Cambridge, MA 02138, USA\\
$^{2}$Department of Astronomy, Columbia University, 550 West 120th Street, New York, NY 10027\\
$^{3}$Einstein Fellow}

\begin{document}
\maketitle

\begin{abstract}
We study the interaction of a supermassive black hole (SMBH) binary
and a standard radiatively efficient thin accretion disk.  We examine
steady-state configurations of the disk and migrating SMBH system,
self-consistently accounting for tidal and viscous torques and
heating, radiative diffusion limited cooling, gas and radiation
pressure, and the decay of the binary's orbit.  We obtain a ``phase
diagram'' of the system as a function of binary parameters, showing
regimes in which both the disk structure and migration have a
different character.  Although massive binaries can create a central
gap in the disk at large radii, the tidal barrier of the secondary
causes a significant pile--up of gas outside of its orbit, which can
lead to the closing of the gap.  We find that this spillover occurs at
an orbital separation as large as $\sim 200 M_7^{-1/2}$ gravitational
radii, where $M_{\SMBH} = 10^7 M_7 \Msun$ is the total binary mass.
If the secondary is less massive than $\sim 10^6\Msun$, then the gap
is closed before gravitational waves (GWs) start dominating the
orbital decay.  In this regime, the disk is still strongly perturbed,
but the piled-up gas continuously overflows as in a porous dam, and
crosses inside the secondary's orbit.  The corresponding migration
rate, which we label Type 1.5, is slower than the usual limiting cases
known as Type I and II migration.  Compared to an unperturbed disk,
the steady-state disk in the overflowing regime is up to several
hundred times brighter in the optical bands.  Surveys such as
PanSTARRS or LSST may discover the periodic variability of this
population of binaries.  Our results imply that the circumbinary disks
around SMBHs can extend to small radii during the last stages of their
merger, when they are detectable by {\it LISA}, and may produce coincident
electromagnetic (EM) emission similar to active galactic nuclei (AGN).
\end{abstract}

 \begin{keywords}
accretion, accretion discs -- black hole physics -- gravitational waves -- galaxies: active
 \end{keywords}

%
%
%
%

\section{Introduction}
\subsection{Overview}

Understanding the coevolution of binaries and the accretion disks in
which they are embedded is critical in several fields of astrophysics,
including planet formation and migration
\citep{1980ApJ...241..425G,1997Icar..126..261W}, patterns in planetary
rings \citep{1982ARA&A..20..249G}, stellar binaries \citep{1987ARA&A..25...23S,2007ARA&A..45..565M},
the final parsec phase of black hole binaries
\citep{1980Natur.287..307B,2005ApJ...630..152E,2009MNRAS.398.1392L},
stars and black holes (BHs) in active galactic nuclei
\citep{2004ApJ...608..108G,2005ApJ...619...30M,2007MNRAS.374..515L},
and electromagnetic (EM) counterparts to gravitational wave (GW)
events
\citep{2008ApJ...684..870K,2009CQGra..26i4032H,2011CQGra..28i4021S}.
These systems may also provide observational probes to test models of
the anomalous viscosity in accretion disks
\citep{2011PhRvD..84b4032K,2011PhRvL.107q1103Y}.

Despite the long history of the subject, there are very few
self-consistent analytic
models for the coevolution of binaries and accretion disks,
incorporating the fundamental physical effects over the long
timescales on which the binary separation evolves.  The standard
$\alpha$--model of radiatively efficient turbulent thin accretion
disks \citep{1973A&A....24..337S} relates the effective kinematic
viscosity of the disk to the pressure $\nu \propto \alpha p$.  The viscous
evolution of the disk, however, is often modeled without considering
the pressure dependence of the viscosity
\citep{1974MNRAS.168..603L}. Similarly, models of the gravitational
interaction between the disk, which describe the launching of spiral
density waves in the disk that remove angular momentum from the
binary, also do not account for the tidal heating of the disk and the
corresponding feedback on the torques \citep{1980ApJ...241..425G}.

In an accompanying paper \citep[hereafter Paper~I]{paper1} we derive
an analytic steady-state model for the coevolution of the disk and
the orbital migration of the secondary, in which we combine a
\citet{1973A&A....24..337S} disk with the theory of the binary-disk
interaction by \citet{1980ApJ...241..425G} self-consistently.  In
particular, we adopt the viscosity prescription for standard thin
$\alpha$ or $\beta$ disks, calculate the sound speed and vertical
balance including both gas and radiation pressure ($p_{\rm gas}$ and
$p_{\rm rad}$), adopt a simple analytic approximation to the angular
momentum exchange between the binary and the disk of
\citet{2002ApJ...567L...9A}, consider the viscous and tidal heating of
the disk \citep{2009MNRAS.398.1392L}, and self-consistently account
for the feedback on the pressure, viscosity, scaleheight, and the
torque cutoff near the secondary's orbit, as well as on the migration
rate of the secondary.  We derive azimuthally averaged analytic disk
models which recover the \citet{2004ApJ...608..108G} solution for
arbitrary $\beta=p_{\rm gas}/(p_{\rm gas}+p_{\rm rad})$ in the limit
that the secondary mass $m_{\SCO}$ approaches zero.  The solution is
then generalized for larger $m_{\SCO}$, i.e. to the case when the disk
structure is significantly modified by the secondary, over multiple
accretion timescales.

In this paper, we explore the implications of the new binary+disk
evolution solutions, found in Paper~I, for SMBH binary systems.  By
varying the binary parameters systematically, we explore possible
distinct behaviors of the disk--secondary system.  We recover the two
limiting cases known previously, and identify a new intermediate
phase.  Low-mass objects perturb the disk only weakly, and the linear
density perturbations lead to the extensively studied \emph{Type-I
migration} of the secondary \citep{1980ApJ...241..425G}. For very
massive objects, the tidal torque clears a gap in the disk, and the
viscous radial inflow of the gas pushes the object inward (known as
\emph{Type-II migration}). A particular subclass of the latter is the
so-called secondary-dominated Type-II migration, in which the
secondary's mass exceeds the nearby gas mass, causing the migration to
slow down, and the surface density outside of the gap to build up,
before it is able to efficiently push the object inward
\citep{1995MNRAS.277..758S}.  This assumes that the pile-up is $100\%$
efficient and no gas can cross the secondary's orbit.  We identify a
separate, intermediate class of migration, \emph{Type-1.5}, in which
the gas piles up significantly outside of the perturber's orbit, but
the viscosity increases to the point that in steady-state, the gas
enters the nonlinear gravitational field of the secondary (i.e. its
Hill sphere), and is able to flow across its orbit. Not surprisingly,
the corresponding migration rate is significantly different from both
the Type I and II cases.\footnote{{
The non-axisymmetric part of the perturbations
may} {be either linear or non-linear depending on the
magnitude of pile-up.
In both cases, we adopt the tidal torque model of \citet{2002ApJ...567L...9A},
reminiscent of Type-I migration, which we use with the
azimuthally--averaged self-consistent disk profile (see Paper~I).
}}

In most previous investigations, the gap opening conditions were based
on the comparison of viscous and tidal torques in a weakly perturbed
disk
\citep{2006Icar..181..587C,2007astro.ph..1485A,2011PhRvD..84b4032K}.
The orbital decay of the binary and the evolution of the disk density
profile is then coupled; this coupled evolution has not been followed
on the long time-scales on which the orbit evolves.  Although a cavity
may indeed be opened at large radii, in accretion disks where the gas
build-up outside the cavity is significant, the cavity may close after
several accretion timescales.  Our solutions allow us to derive the
long-term gap opening and closing criteria.

There are many possible observational implications of our findings for
black hole binaries, and perhaps also for planetary dynamics. Here we
restrict our attention to the former context.  First, gap-closing
makes it more likely that GW inspiral events are accompanied by EM
emission, since the binary is embedded in a gaseous disk, with no
central cavity, even at the last stages of the merger.  Previously,
\citet{Liu+2003} and \citet{2005ApJ...622L..93M} have argued that as
the GW inspiral accelerates beyond the rate at which the gas at the
edge of the cavity can viscously follow the binary, the binary
decouples from the disk.  Consequently, they argued that luminous AGN
(active galactic nuclei) or radio emission from jets are expected only
after the gas has had time to accrete onto the remnant.  The
post-merger delay is between years and decades for binaries in the
mass range $10^5$--$10^6\Msun$ expected to be detectable by
{\it eLISA/NGO}\footnote{\url{http://elisa-ngo.org/}}
(\citealt{2012arXiv1201.3621A}; see also
\citealt{2010AJ....140..642T,2010ApJ...714..404T,2010PhRvD..81b4019S}).  However, if the
central cavity refills before the GW emission becomes significant
(which we find is the case in the above mass range, in particular),
then the gas can accrete onto the primary and shine much like a normal
bright AGN, even during the last stages of the merger, producing
coincident EM counterparts or precursors to {\it LISA} sources.

Second, gas accumulation outside of the secondary leads to a greatly
enhanced surface brightness \citep{2009MNRAS.398.1392L}.  This may
help in searches for EM counterparts to more massive
($10^8$--$10^9\Msun$) SMBH binaries at larger separations, still in
the gas-driven stage, emitting GWs in the pulsar timing array
frequency bands \citep{2012MNRAS.420..705T,2012MNRAS.420..860S}. The
EM spectrum of the disk missing the emission at high frequencies if
the disk has a gap \citep{1995MNRAS.277..758S}, and the bright gap
edge has a characteristic ultraviolet-optical-infrared profile
\citep{2009MNRAS.398.1392L}.  We investigate the brightening as a
function of component masses and separations.  If these sources
produce correspondingly bright periodic EM variability on the orbital
timescale, they can be identified in future time-domain optical/IR
surveys.  The statistics of many such sources can observationally test
the migration and GW inspiral rates \citep{2009ApJ...700.1952H}.
Combining the predictions for the variability timescale, disk
brightness, and spectrum offers new independent tests of the physical
models of the accretion physics, disk-satellite interactions, and GW
emission.

{\citet{2004ApJ...608..108G} and \citet{2012arXiv1206.2309M} pointed out that
supermassive stars or intermediate mass black holes (IMBHs) in the range $10^2$--$10^5\,\Msun$
may form in AGN disks. These objects would perturb the accretion disk, 
causing a pile up and gap overflow as described here. Our studies 
imply a slow-down of migration for these objects, 
making these systems longer-lived, and thus increasing the likelihood of their detection.}

\subsection{Relation to previous works}\label{s:literature}

Without the aim of completeness, we highlight here the similarities
and main differences between our study and some related papers in the
recent literature.

Based on the solutions of \citet{1967pswh.book.....Z} and \citet{1991MNRAS.248..754P},
\citet{1999MNRAS.307...79I} showed that the 
circumbinary disk with a binary evolves in a self-similar way on scales much 
larger than the binary, assuming that 
the initial and outer boundary conditions represent an unperturbed 
stationary $\alpha$-disk with a fixed $\dot{M}$, and assuming that
a gap is always present which truncates the disk within the secondary 
and causes the radial gas velocity to be effectively zero near the secondary
(see also \citealt{2012arXiv1205.5017R}). 
Here, we examine the opposite limiting case in which the 
radial gas velocity is nonnegligible near the secondary due to gap overflow
and the disk is approximately in a steady-state.

\citet{2010MNRAS.407.2007C} examined the coevolution of the disk and
the secondary, and found that the former may exhibit a rapid
brightening in the GW driven regime as the binary shepherds gas inward
before merger (however, see \citealt{2012MNRAS.tmpL.436B}). 
They included almost the same physics as this study and
solved the time dependent equations in 1 dimension numerically for
a binary with $10^7$ and $10^6\Msun$ mass components. Their initial conditions were that of an 
unperturbed disk and focussed on the final GW driven regime. In this
case, they found no gas overflow across the gap. 

\citet{2010PhRvD..82l3011L} constructed an analytic steady-state
model, and showed that the disk may brighten significantly when an
object is placed in the disk (locally by a factor of $10^4$ for a mass
ratio $q=0.1$).  However, they have considered a constant $H/r$ (here
$H$ is the disk scaleheight and $r$ is the radius) and a viscosity
profile corresponding to an unperturbed disk, and neglected the
changes in these quantities due to the secondary.

\citet{2009MNRAS.398.1392L} considered a disk model whose local
physics is very similar to ours, but solved the time evolution
numerically in 1 dimension for a very different choice of initial and
boundary conditions.  In particular, they focused on the specific case
where the SMBH binary is very massive, and has a ``one-time'' disk
that is much less massive, compact (spreading over at most a factor of
10 in radii)
and is not replenished by accreting new material from large radii.
They found gas pileup outside the secondary, leading to a brightening
of the outer disk, and a modified spectrum.  Their numerical results
provide a useful independent reference to qualitatively verify our
steady-state radial disk surface density and scaleheight profiles.
For the particular binaries they have considered, the migration rate
was greatly reduced, such that the binary is not transported to the
GW-driven regime within a Hubble time, and they showed that this poses
an obstacle against solving the final parsec problem
\citep{1980Natur.287..307B}.  In our paper, we examine the
steady-state configuration, under the assumption of a constant
accretion rate in the disk (set by the Eddington limit near the
primary).  This assumes a constant mass supply from larger radii, and
is therefore very different from the ``one-time'' disk in
\citet{2009MNRAS.398.1392L}.
As a result, we reach essentially the opposite conclusions: we find a
stronger pileup, which increases radiation pressure in the disk and
stabilizes it against gravitational fragmentation, and yields a much
faster migration (well within the Hubble time).  Another practical
difference in our study is that we consider a broad range of binary
masses, mass ratios and semimajor axes, including the radiation
pressure dominated regime, and we map out the distinct phases for disk
structure and migration.

Most studies on planetary migration neglect the tidal heating effect
and radiation pressure, and are therefore inapplicable for our
purposes.  We investigate AGN accretion disks where radiation effects
are more significant. Interestingly, we find that the disk becomes
strongly radiation pressure dominated outside the secondary, even in
regions far from the primary, which, in the absence of the secondary,
would be gas pressure dominated.  Therefore, this requires treating
the fluid as comprised of both gas and radiation.  The use of a single
equation of state parameter, as in most papers in planetary dynamics,
becomes invalid in this regime.  We note that
\citet{2003ApJ...599..548D} did account for tidal heating and
temperature variations in a two-dimensional simulation, but neglected
the effects of radiation pressure. {\citet{2006A&A...459L..17P,2008A&A...478..245P} and
\citet{2008A&A...487L...9K}}
presented results from numerical simulations with radiation,
showing that tidal heating and radiation pressure have a significant
effect on the migration of planets.

Criteria for gap opening and closing have been investigated
extensively for protoplanetary disks
\citep{1986ApJ...309..846L,1994ApJ...421..651A,1997Icar..126..261W,2006Icar..181..587C}.
Previous numerical studies typically neglected the effects of gas
build-up outside the gap, and did not consider self-consistently the
effects of the excess viscous and tidal heating of the gas, and
neglected both radiation pressure and the migration of the secondary.
However, two- and three-dimensional simulations have shown that even
if the gap opening conditions are satisfied, gas can periodically flow
in along non-axisymmetric steams into the gap and accrete onto the
primary and the planet, particularly if the ratio of the distance to
the gap edge to the Hill radius is of order unity
\citep{1996ApJ...467L..77A,1999ApJ...526.1001L,2006ApJ...641..526L,Hayasaki+07,2008ApJ...672...83M,2009MNRAS.393.1423C}.
In this paper, we revisit the standard gap opening/closing conditions
in circumbinary accretion disks, including the effects of radiation
pressure \citep{2011PhRvD..84b4032K}, as well as gas build-up, tidal
heating, and migration. For simplicity, we neglect a possible
non-axisymmetric inflow if the gap is larger than the Roche lobe and
also neglect accretion onto the secondary. We also do not model the
magneto-rotational instability (MRI), which may influence the conditions
for gap opening \citep{2003ApJ...589..543W,2012arXiv1204.1073N,2012ApJ...749..118S,2012arXiv1207.3354F}.
These effects are likely to
be important, and should be investigated in the future in two- or
three-dimensional simulations.

\subsection{Outline and conventions}
This paper is organized as follows.
In \S~\ref{s:interaction}, we briefly lay out the physical model that
we adopt.
In \S~\ref{s:disk}, we present numerical and analytic solutions to
the disk and identify its distinct physical phases.
In \S~\ref{s:gap}, we elaborate on the gap opening and closing
conditions in steady-state.
In \S~\ref{s:migration}, we discuss the migration of the secondary,
comparing the new Type-1.5 solution in a continuously overflowing disk
with the Type-I and Type-II cases.
In \S~\ref{s:observations}, we discuss observable implications,
including the lightcurve, spectrum, the abundance of AGN with periodic
variability, as well as GW observations with {\it LISA/NGO} and pulsar timing
arrays (PTAs, \citealt{2010CQGra..27h4013H}).  We summarize our main
conclusions in \S~\ref{s:conclusions}.  The interested reader can find
details of analytic derivations in Paper~I.

\begin{figure}
\centering
\mbox{\includegraphics[width=8.5cm]{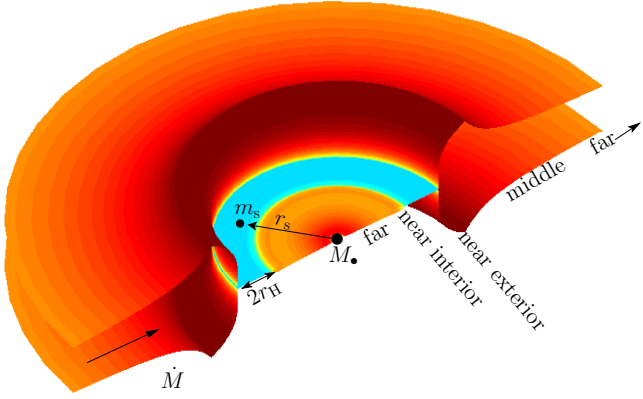}}
\caption{\label{f:diskpic}
Gas pile-up and overflow in a circumbinary accretion disk with
component masses $M_{\SMBH}$ and $m_{\SCO}$, binary separation
$r_{\SCO}$, and accretion rate $\dot{M}$. We distinguish five distinct
radial zones in the disk: an inner and an outer far zone where the
effects of the secondary are negligible, an interior and an exterior
near zone, where the tidal effects are significant, and an extended
middle zone, where the disk is still significantly perturbed. Our
model breaks down inside the secondary's Hill radius, denoted by
$r_{\rm H}$.}
\end{figure}

Our basic notation for the disk and secondary parameters are depicted
in Figure~\ref{f:diskpic}.  We denote the primary and secondary mass with
$M_{\SMBH}$ and $m_{\SCO}$, and the mass ratio with $q \equiv
m_{\SCO}/M_{\SMBH}$.  We use geometrical units $\G=\C=1$ and suppress
factors of $\G/\C^2$ and $\G/\C^3$ in conversions between mass,
length, and time units. We use $\R\equiv r/(\G M_{\SMBH}/\C^2)$ to
label the radius in gravitational radii.  A subscript ``$\SCO$''
refers to quantities describing the secondary, while the subscript
``0'' refers to quantities in the unperturbed disk around a single
compact object.

\section{Thermo-hydrodynamical interaction between a disk and a secondary}\label{s:interaction}

We examine the evolution of the secondary and an azimuthally and
vertically averaged Shakura-Sunyaev disk (i.e. axisymmetric one-zone
disk) in local thermal equilibrium.  The equations are laid out in
detail, and solved both numerically and under various approximations,
analytically in Paper~I.  Here we only briefly summarize the basic
conceptual framework and the features of the solutions; we refer the
reader to Paper~I for the complete set of equations and a detailed
description.


\subsection{Equations governing the evolution of the disk structure and the binary's orbit}

We denote the azimuthally-averaged surface density profile of the disk by $\Sigma(r)$,
and the radial (bulk) velocity of the gas in the disk by
$v_r(r)=-\dot{M}/[2\pi r \Sigma(r)]$, which is negative as gas
accretes toward the primary at $r=0$ with an accretion rate $\dot{M}$.
We assume a nearly Keplerian disk and denote the angular velocity with
$\Omega$.

Assuming that the migration velocity of the secondary, $v_{\SCO r}$ is
sufficiently slow, the disk evolves through a sequence of steady-state
configurations, where $\dot{M}(r)=\dot{M}$ is a constant.  The angular
momentum flow in the disk is driven by the viscous torques ($T_{\nu}$)
and the tidal torque exerted by the secondary ($T_{\rm d}$), while the
angular momentum loss of the secondary is due to the backreaction of
the tidal torque and gravitational wave losses
\begin{align}\label{e:Tnu'}
\dot{M}\partial_r(r^2 \Omega) &=  \partial_r T_{\nu} - \partial_r T_{\rm d}\,,\\
\label{e:SCO}
\dot{L}_{\SCO} &= \frac{1}{2} m_{\SCO} r_s \Omega_s v_{\SCO r} = -\int_0^{\infty} \partial_r T_{\rm d}\; \D r - T_{\GW}\,,
\end{align}
Here
\begin{align}
T_\nu &=  -2 \pi r^3 (\partial_r \Omega)\,\nu \Sigma \simeq 3\pi\, r^2 \Omega\, \nu \Sigma \,, \label{e:Tnu}\\
\partial_r T_{\rm d} &= 2\pi r  \Lambda \Sigma \,,\label{e:Td}\\
T_{\rm GW} &= \frac{32}{5} \frac{m_{\SCO}^2}{M_{\SMBH}} \R_{\SCO}^{-7/2}\,,
\end{align}
where we adopt the widely-used approximation of
\citet{2002ApJ...567L...9A} for the specific tidal torque,
\be\label{e:Lambda}
\Lambda \approx
\left\{
\begin{array}{c}
 - \frac{1}{2} f q^2 r^2\Omega^2  r^4/\Delta^4 \text{~~if~} r < r_{\SCO}-r_{\rm H}\,,\\
 +\frac{1}{2} f q^2 r^2\Omega^2 r_{\SCO}^4/\Delta^4\text{~~if~} r > r_{\SCO}+r_{\rm H}\,,
\end{array}
\right.
\ee
where
\begin{equation}\label{e:Delta}
\Delta \equiv \max(|r - r_s|, H)\,.
\end{equation}
Equation~(\ref{e:Lambda}) breaks down in the region near the
secondary's orbit, i.e. within its Hill radius or tidal radius, $|r-r_{\SCO}| < r_{\rm
H} \equiv (q/3)^{1/3} r_{\SCO}$. We excise this region from our
calculations and instead match the interior and exterior solutions, setting
$T_{\nu}(r_{\SCO}-r_{\rm H}) = T_{\nu}(r_{\SCO}+r_{\rm H})$.
This has an effect similar to smoothing the torques inside the Hill
radius, as done previously in \citet{1986ApJ...309..846L},
\citet{1995MNRAS.277..758S}, and \citet{2009MNRAS.398.1392L}.  Here,
$q \equiv m_{\SCO}/M_{\SMBH}$, $H\ll r$ is the scaleheight of the
disk, and $f$ is a constant calibrated with simulations.
We conservatively adopt the
low value $f_{-2}\equiv f/10^{-2}=1$ for our numerical solutions, but
the dependence on $f_{-2}$ is explicitly computed in the analytic
solutions.

Given the viscosity $\nu(r)$ and the scale-height $H(r)$,
Eq.~(\ref{e:Tnu'}) can be integrated to find the surface density of
the disk, $\Sigma(r)$, for any location of the secondary, and
Eq.~(\ref{e:SCO}) gives the inward migration velocity of the
secondary \citep{2010PhRvD..82l3011L}.  However $\nu(r)$ and $H(r)$ are not known apriori, since
they depend on $\Sigma(r)$ and the mid-plane temperature $T_{\rm
  c}(r)$. {with the standard Shakura-Sunyaev ansatz $\nu(r) = \alpha c_{s}(r)
H(r)$ for viscosity or its variant where the viscosity is assumed to scale with
the gas (rather than the total) pressure, $\nu(r) = \alpha c_{s}(r)
H(r)\beta(r)$, where $\beta(r)\equiv p_{\rm gas}(r)/p(r)$ (also known
as a ``$\beta$-disk'').  Since $\alpha$ disks are thought to suffer
from a thermal instability (but see \citealt{2009ApJ...691...16H}), in the solutions discussed in this paper,
we}  {follow earlier work and focus on the latter case (also known as a
``$\beta$-disk'').  We derive the
temperature profile assuming thermal equilibrium between gas and radiation,
radiative cooling which balances the heating associated with viscous
dissipation tidal heating.}
These provide a non-linear closed set of equations for the
steady-state disk profile. In practice, we note that all of the
equations are nonlinear local algebraic equations at each radius with
the exception of the angular momentum flux equation (\ref{e:Tnu'})
which is a first-order ordinary differential equation for
$T_{\nu}(r)$. We solve these equations for specific boundary
conditions given below, for a fixed value of the secondary orbital
radius. Once the disk profile has been obtained, Eq.~(\ref{e:SCO})
gives the migration velocity of the secondary.


\subsection{Boundary conditions}\label{s:boundary}

We look for steady-state disk solutions with a constant $\dot{M}$,
assuming that the radial profile of the disk relaxes on a time-scale
shorter than the migration time-scale of the secondary. We find the
corresponding equilibrium steady-state configuration of the disk for
each orbital radius of the secondary, and then assume that the disk
proceeds through a sequence of such steady-state configurations as the
secondary migrates inwards.

We distinguish two types of inner boundary conditions, corresponding
to whether or not a gap is assumed to be present in the steady-state
configuration.

{\em I. Boundary condition without a {cavity}.} We here assume that the gas
can continuously overflow, crossing the secondary's orbit, and the
surface density is finite ($\Sigma(r) \neq 0$) throughout the disk all
the way to the primary's innermost stable circular orbit $r_{\rm
ISCO}$ .  We then adopt the zero-torque inner boundary condition
usually assumed in accretion disks onto a single point mass
\citep{1973blho.conf..343N,2010MNRAS.408..752P,2011MNRAS.410.1007T,2012arXiv1202.1530Z},
i.e. we require
\begin{equation}\label{e:TnuISCO}
 T_{\nu}(r_{\rm ISCO})=0\, .
\end{equation}
Starting with this boundary condition, we obtain all properties of the
disk, including the gas velocity profile $v_r(r)$, as well as the
radial speed of the secondary $v_{\SCO r}$. If the disk is strongly
perturbed\footnote{i.e. the viscous torque is greatly reduced (increased) in
the near zone interior (exterior) to the secondary's orbit
relative to the solitary disk without a secondary},
the solution is self-consistent if $|v_r(r)|\gg |v_{\SCO r}|$ over a
wide range of radii around the secondary, so that the disk can relax
sufficiently rapidly to steady-state.  In practice, we assume
approximate steady-state if $v_r(\lambda r_{\SCO})\geq \lambda v_{\SCO
r} $ (where the constant $\lambda\gtrsim 1$ will be introduced below).
Otherwise, the secondary outpaces the nearby gas inflow, and if the
disk is strongly perturbed, we assume a gap opens.  We discuss gap
opening in detail in \S~\ref{s:gap}.

{\em II. Boundary condition {for a truncated disk}.}  If the tidal torques
dominate over the viscous torques near the secondary, gas is expelled
from the region near the secondary and we assume that a circular
cavity forms in the disk.  In the cases we consider, the secondary
mass is sufficiently large that the gas piles up significantly outside
the cavity. The tidal toque acting on the inner edge of the disk has a
sharp cutoff (Eq.~\ref{e:Lambda}). We define a characteristic radius
in the disk outside the gap, where the tidal torque is exerted on the
disk\footnote{In our numerical solutions we define $r_{\rm g}$ as the
radius where the tidal torque density drops to $10\%$ of its peak
value, although our results are insensitive to this precise choice.}
$r_{\rm g} = \lambda r_{\SCO}$ and assume that the density enhancement
at this radius tracks the inward migration of the secondary in a
self-similar way, preserving a constant ratio of radii $ \lambda =
r_{\rm g}/r_{\SCO}$.  This requires that the gas velocity at $r_{\rm
g}$ satisfies
\begin{equation}\label{e:vr0}
 v_r(r_{\rm g})=\frac{r_{\rm g}}{r_{\SCO}} v_{\SCO r}\,.
\end{equation}
Note that $\lambda$ is not specified by hand ab-initio; it is found
self-consistently in our solutions (see below and in Paper~I).
This condition can be understood intuitively, since the secondary
cannot ``run away'' and leave the outer disk behind (if it did, it
would cease to be able to torque the disk and would have to slow
down).  Likewise, the gap edge cannot be moving closer to the
secondary (at least not on on time-scales faster than the migration
timescale; if it did, then the gap would close and the steady-state
solution would be inconsistent).  Although with a moving gap, the disk
cannot strictly be in steady-state near its boundary, we assume
$\dot{M}(r) \approx \dot{M}$ at $r>r_{\rm g}$ (see discussion below).

By construction, only one of the above two boundary conditions will
lead to a self-consistent solution. We speculate that a real
time-dependent binary would evolve through the sequence of
steady-steady solutions we obtain below -- switching between the case
with and without a gap around the transition radii that follows from
the above.

We emphasize that the solutions with a gap are somewhat similar to
those obtained in previous works \citep{1995MNRAS.277..758S}, and also that our solutions in this
regime still suffer from a few possible inconsistencies, as will be
discussed below.  However, the main new result in this paper is the
independent overflowing solution, corresponding to the first of the
two boundary conditions.  As we will argue below, the assumptions
leading to this regime are relatively more robust.  The uncertainties
about the behavior of the disk with a gap could affect only our
results for when the gap closes (as argued below, we took a
conservative approach, in the sense that the ``overflow'' regime may
be present for a wider range of radii than in our fiducial models).

\section{Disk structure}\label{s:disk}
\subsection{Numerical solutions}
\label{s:numerical}

\begin{figure*}
\centering
\mbox{\includegraphics[height=4.3cm]{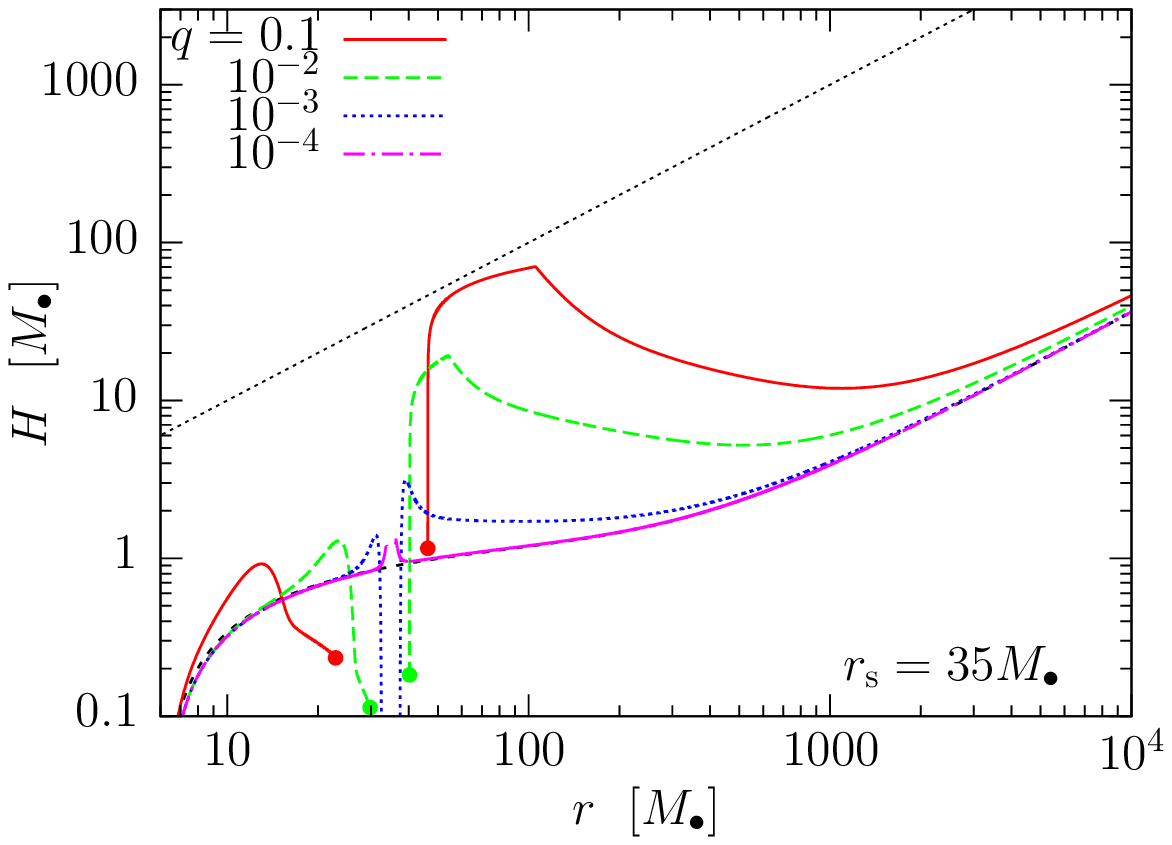}}
\mbox{\includegraphics[height=4.3cm]{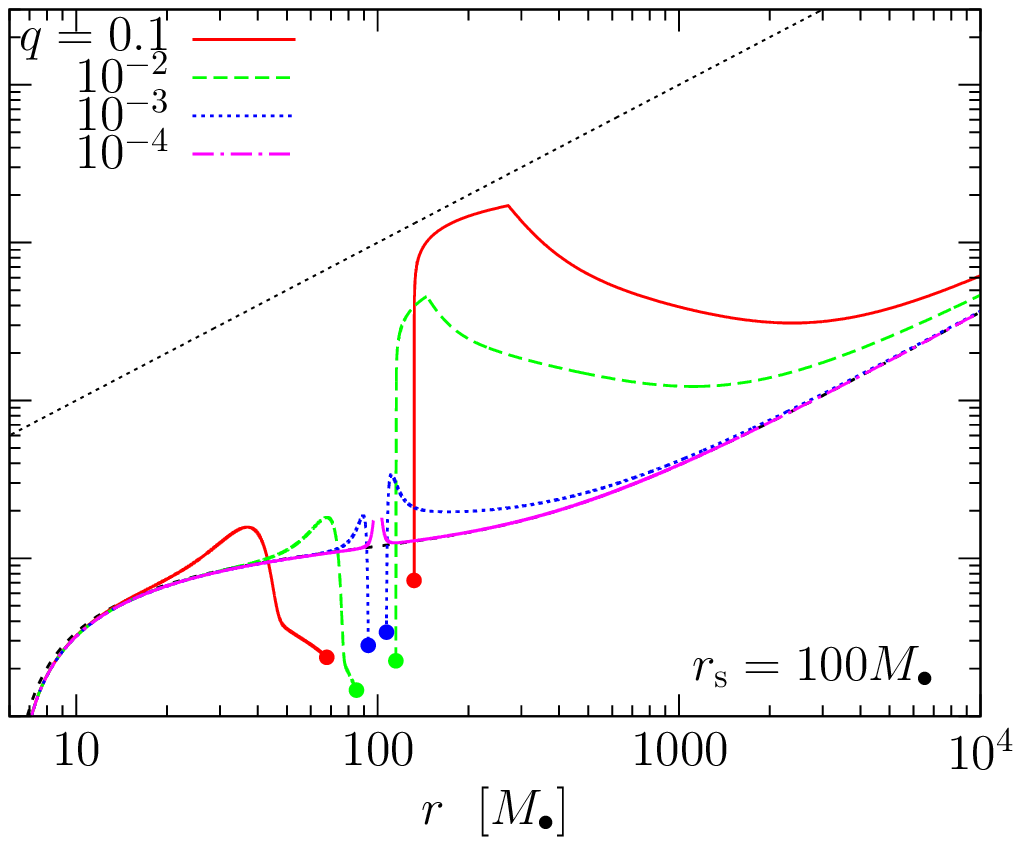}}
\mbox{\includegraphics[height=4.3cm]{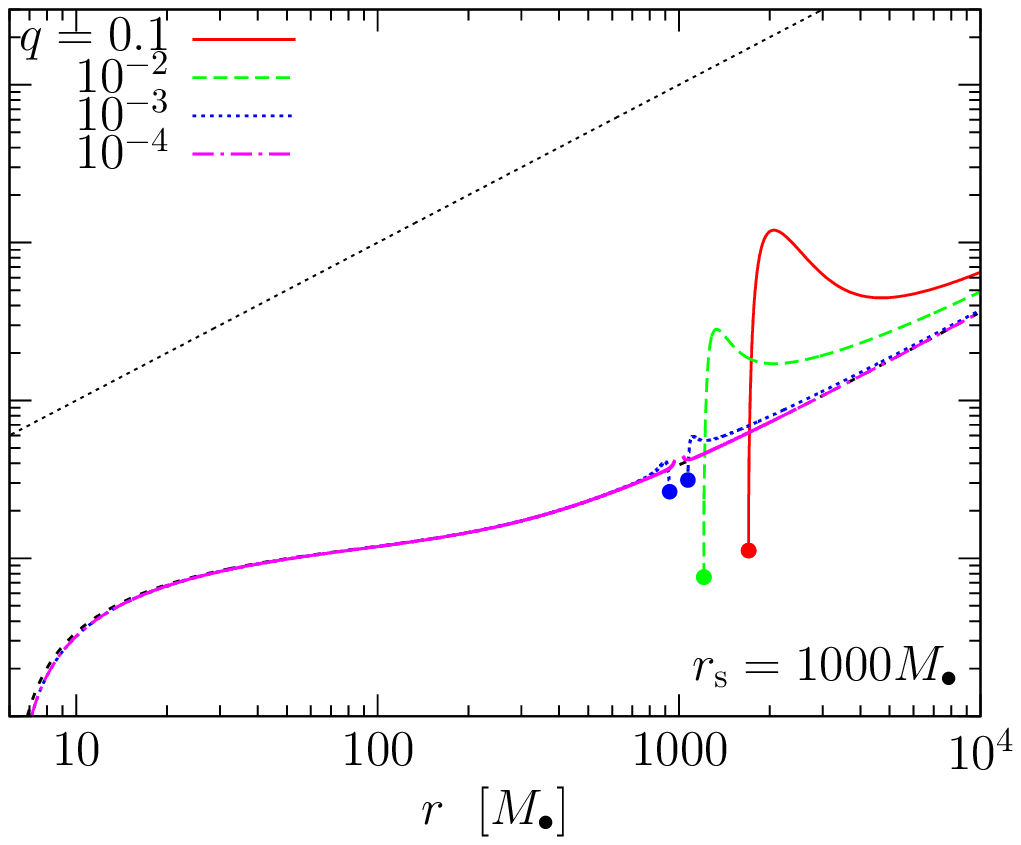}}\\
\mbox{\includegraphics[height=4.3cm]{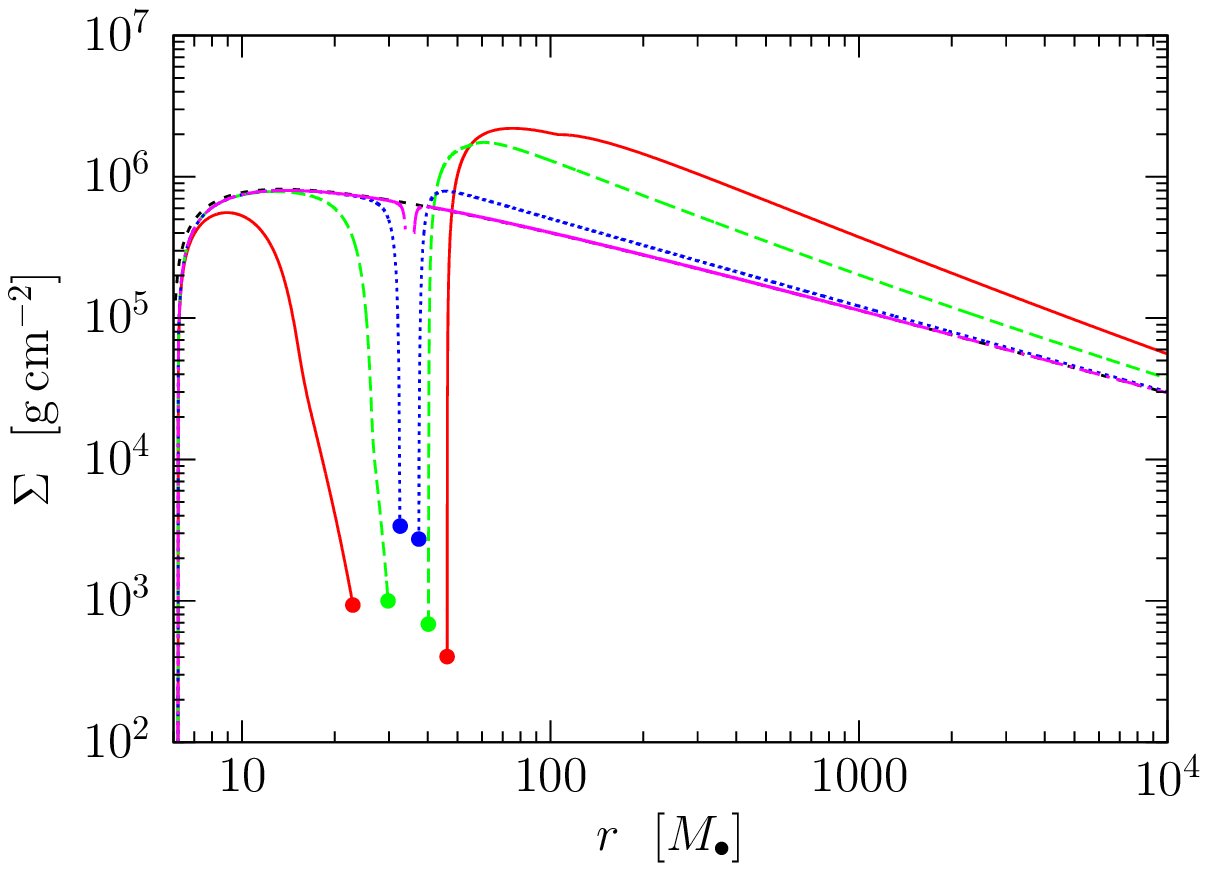}}
\mbox{\includegraphics[height=4.17cm]{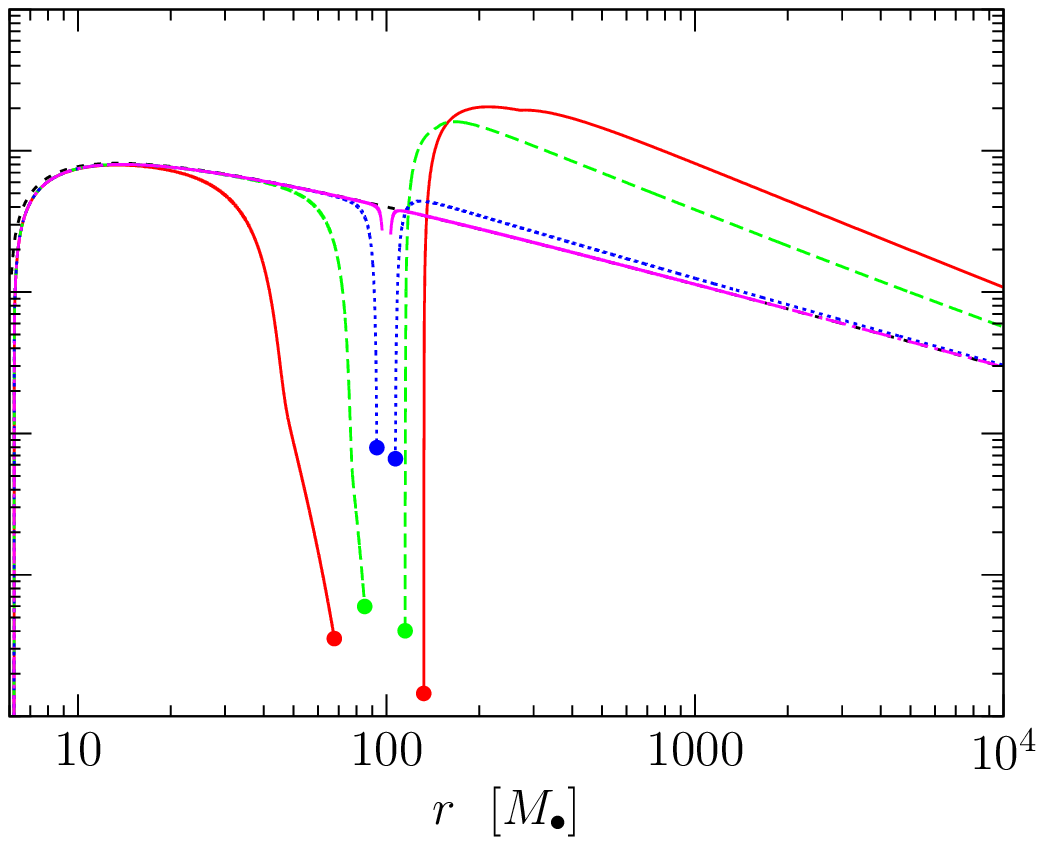}}
\mbox{\includegraphics[height=4.17cm]{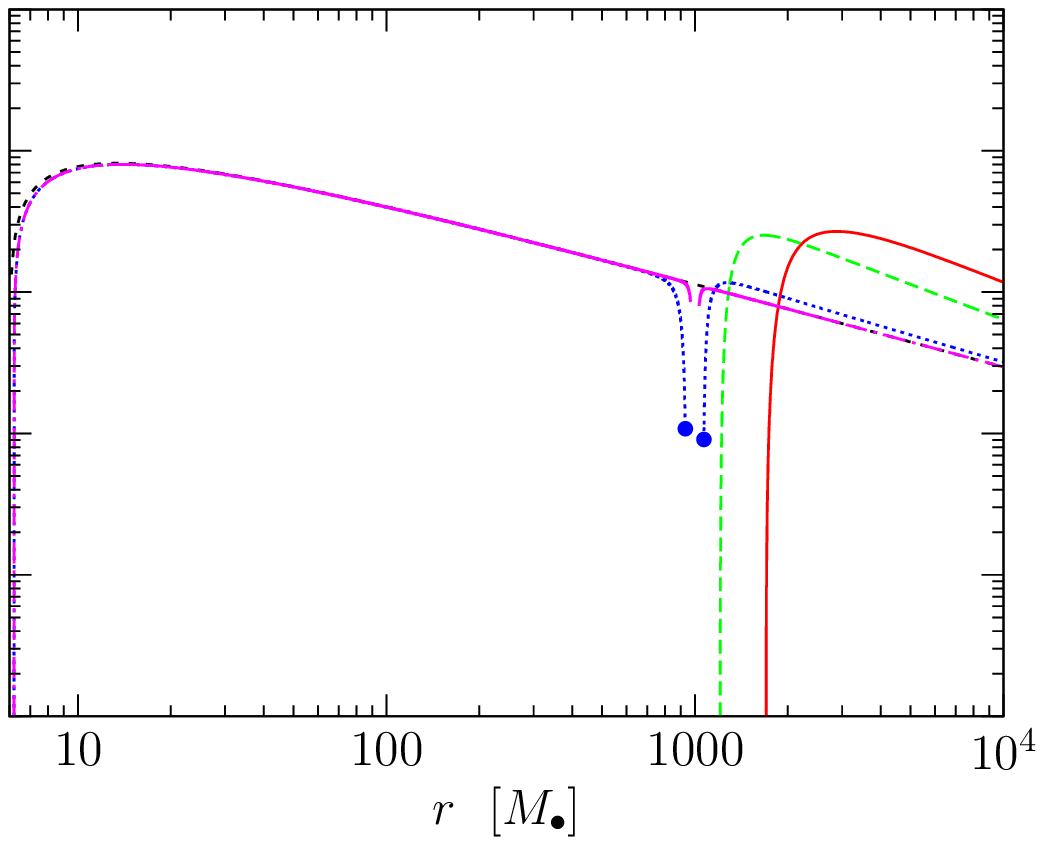}}\\
\mbox{\includegraphics[height=4.3cm]{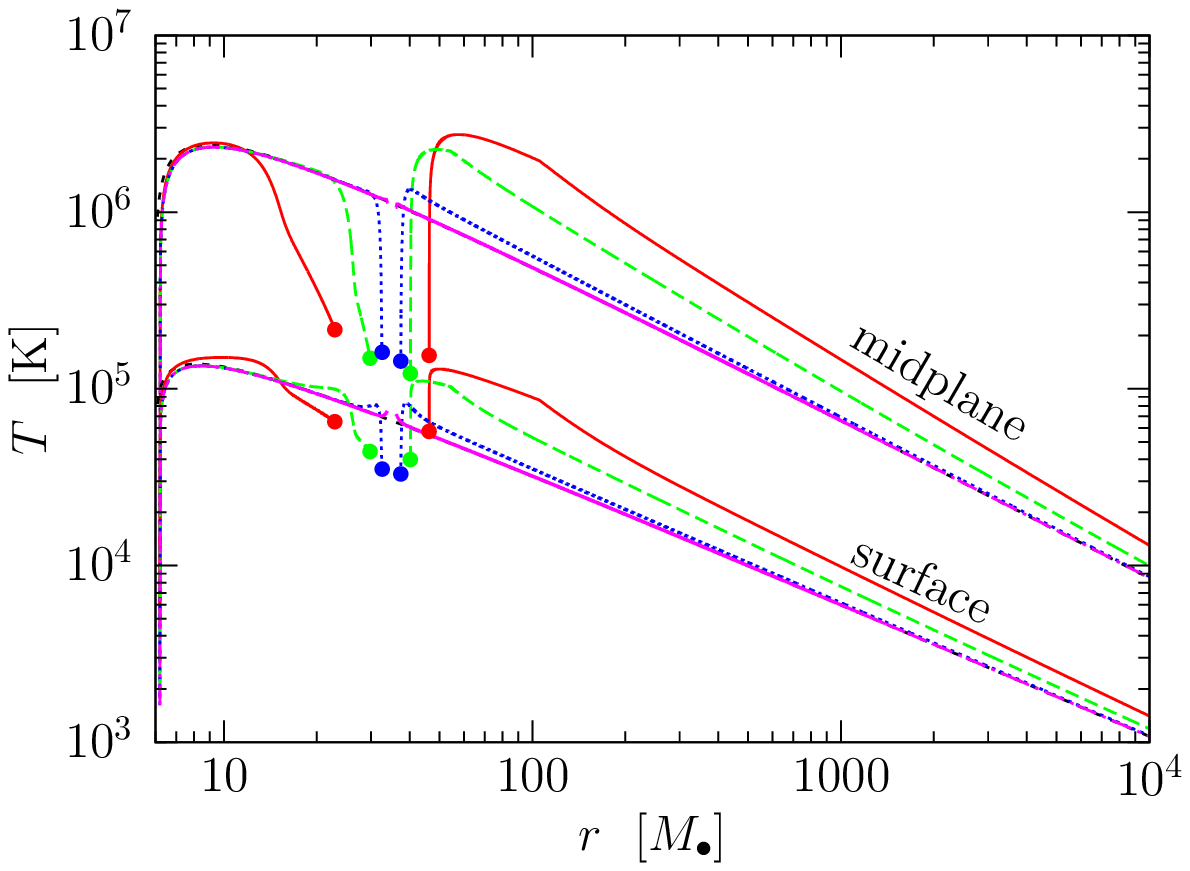}}
\mbox{\includegraphics[height=4.17cm]{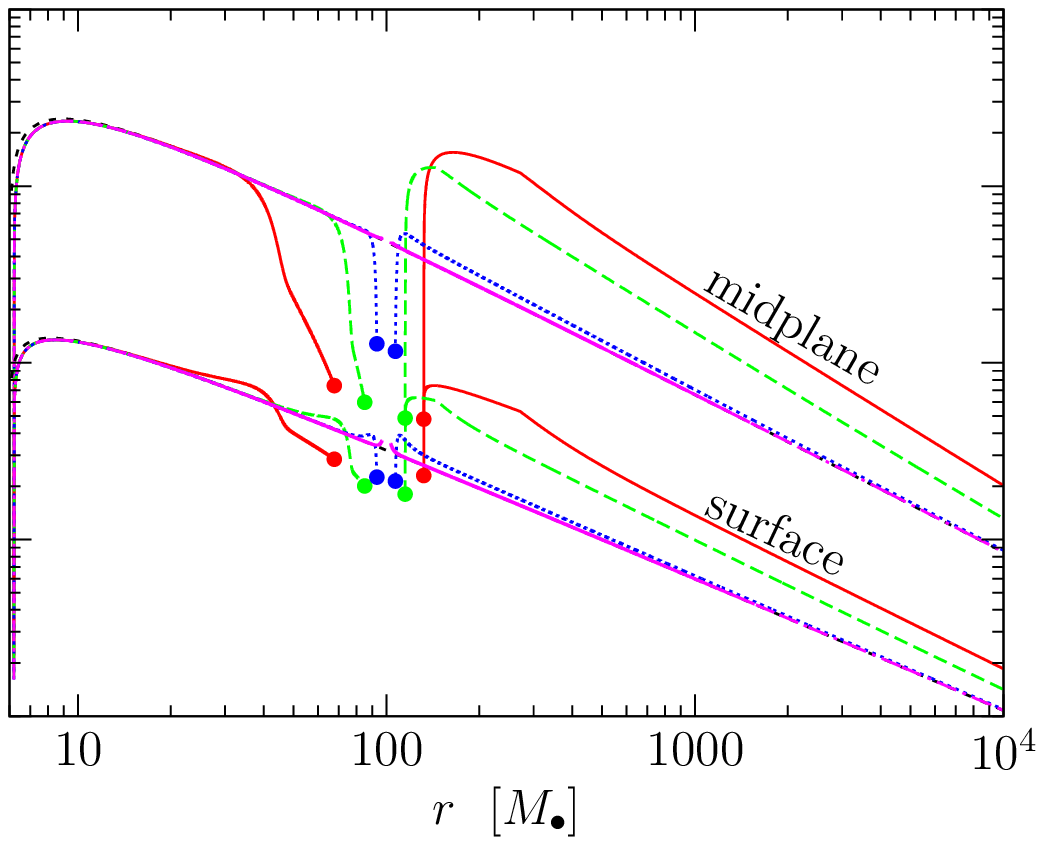}}
\mbox{\includegraphics[height=4.17cm]{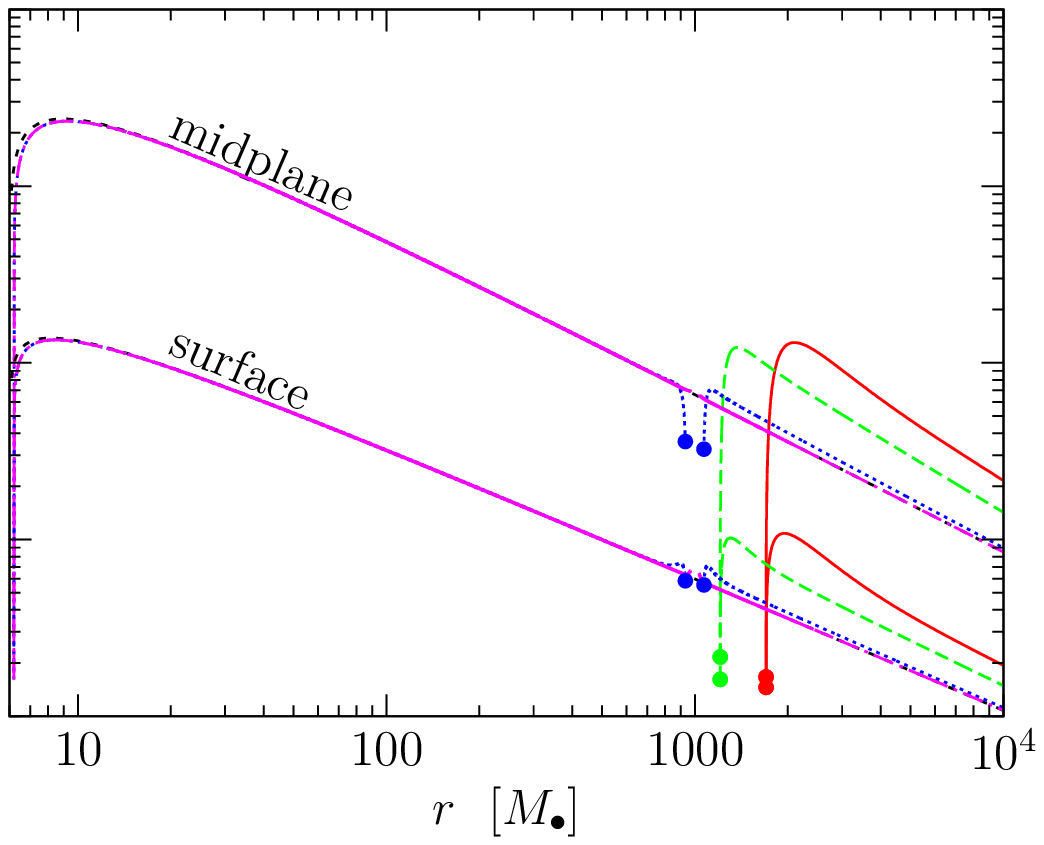}}\\
\mbox{\includegraphics[height=4.3cm]{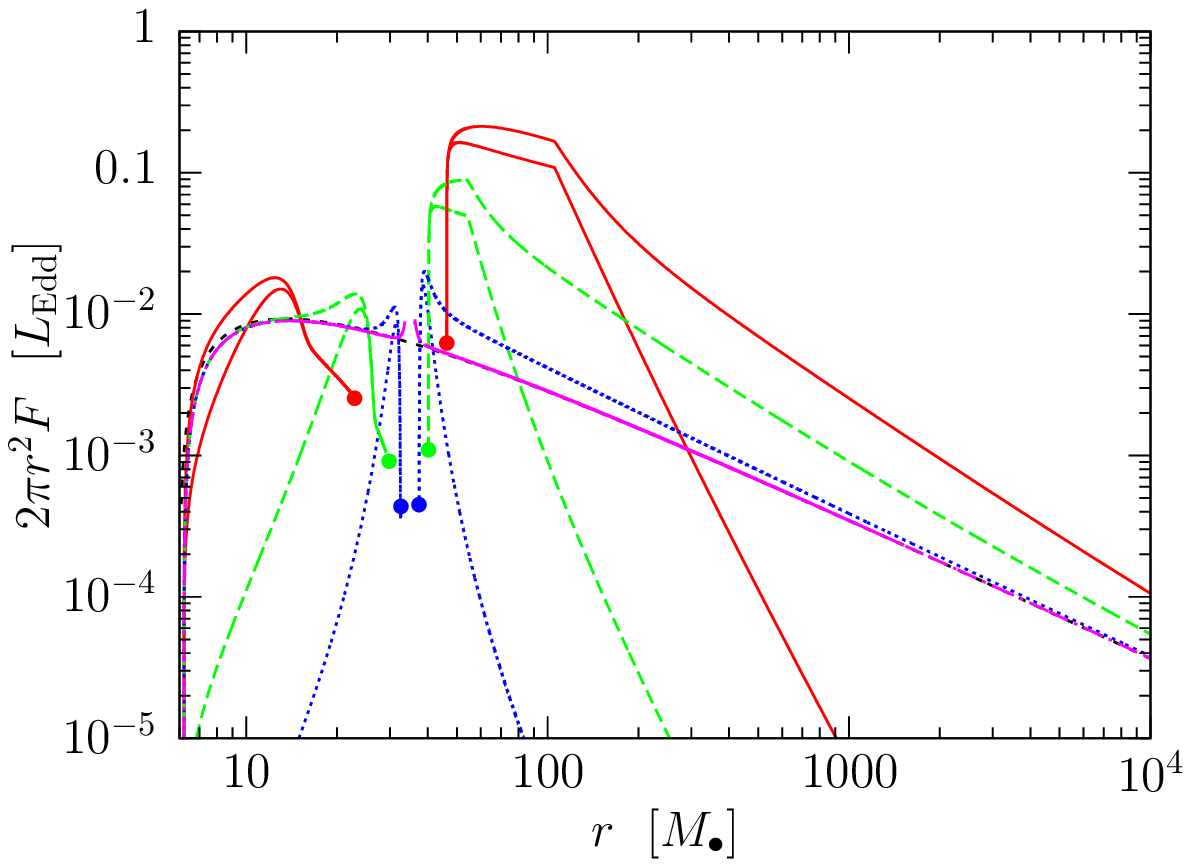}}
\mbox{\includegraphics[height=4.17cm]{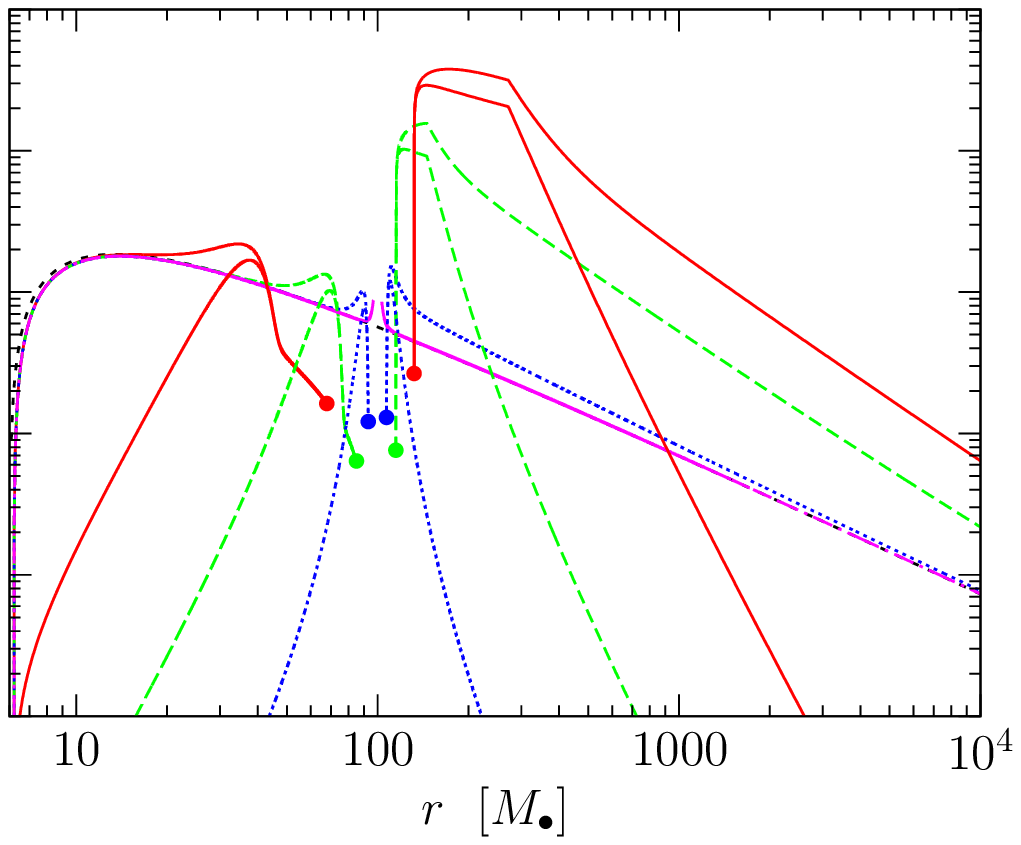}}
\mbox{\includegraphics[height=4.17cm]{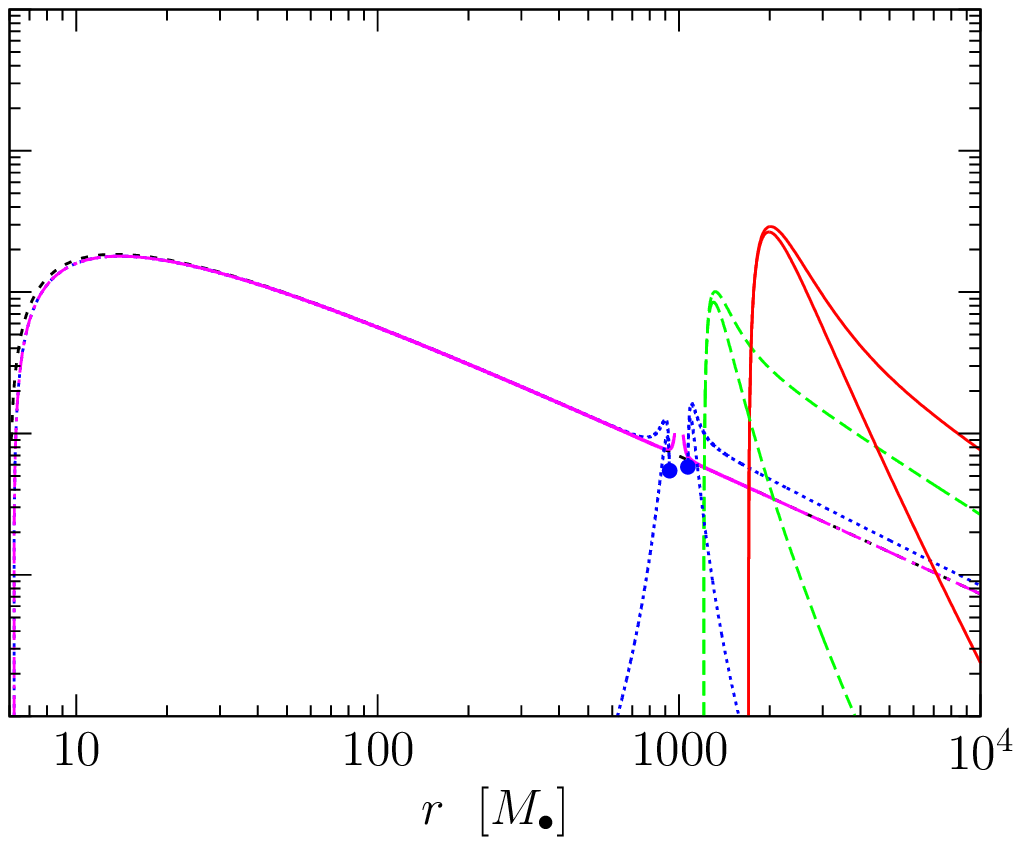}}\\
\mbox{\includegraphics[height=4.3cm]{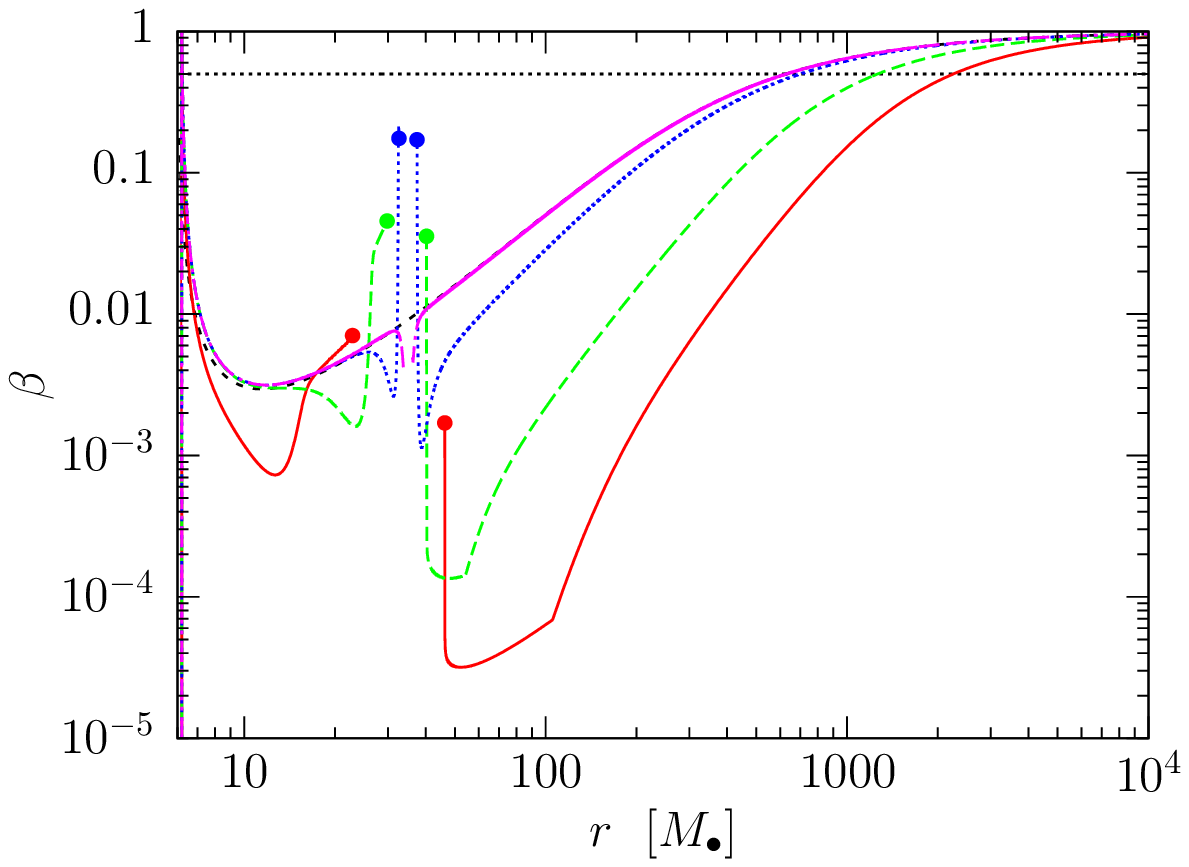}}
\mbox{\includegraphics[height=4.17cm]{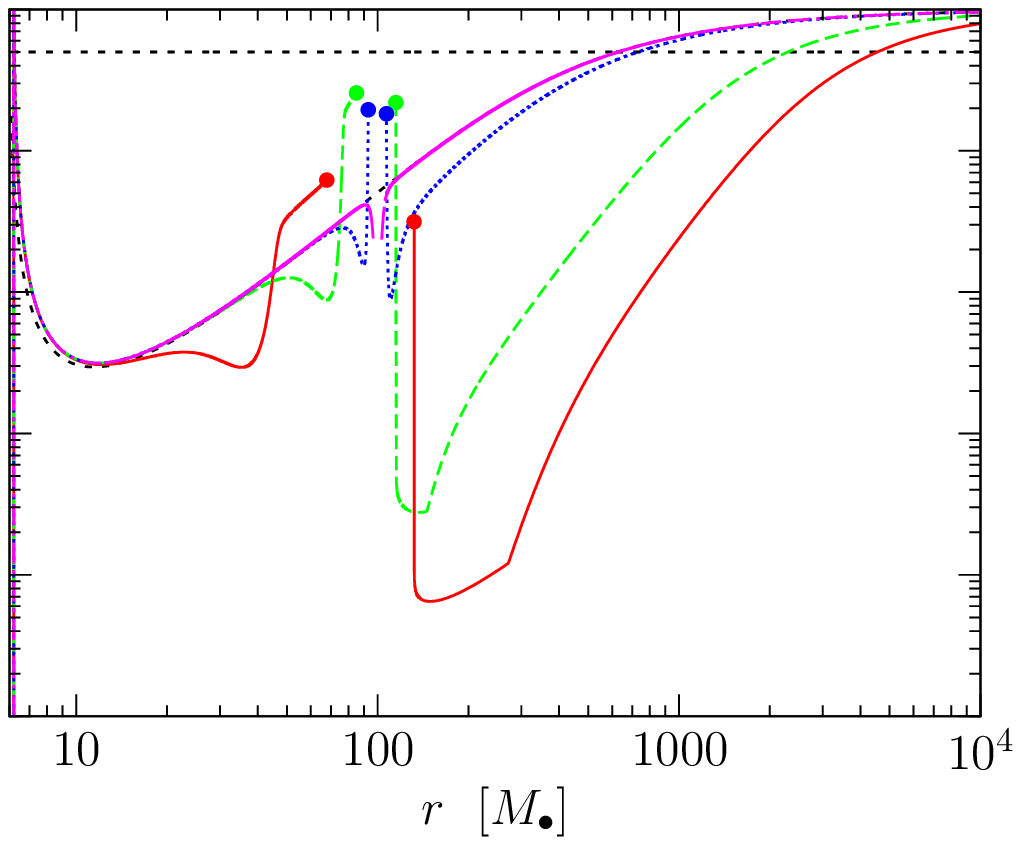}}
\mbox{\includegraphics[height=4.17cm]{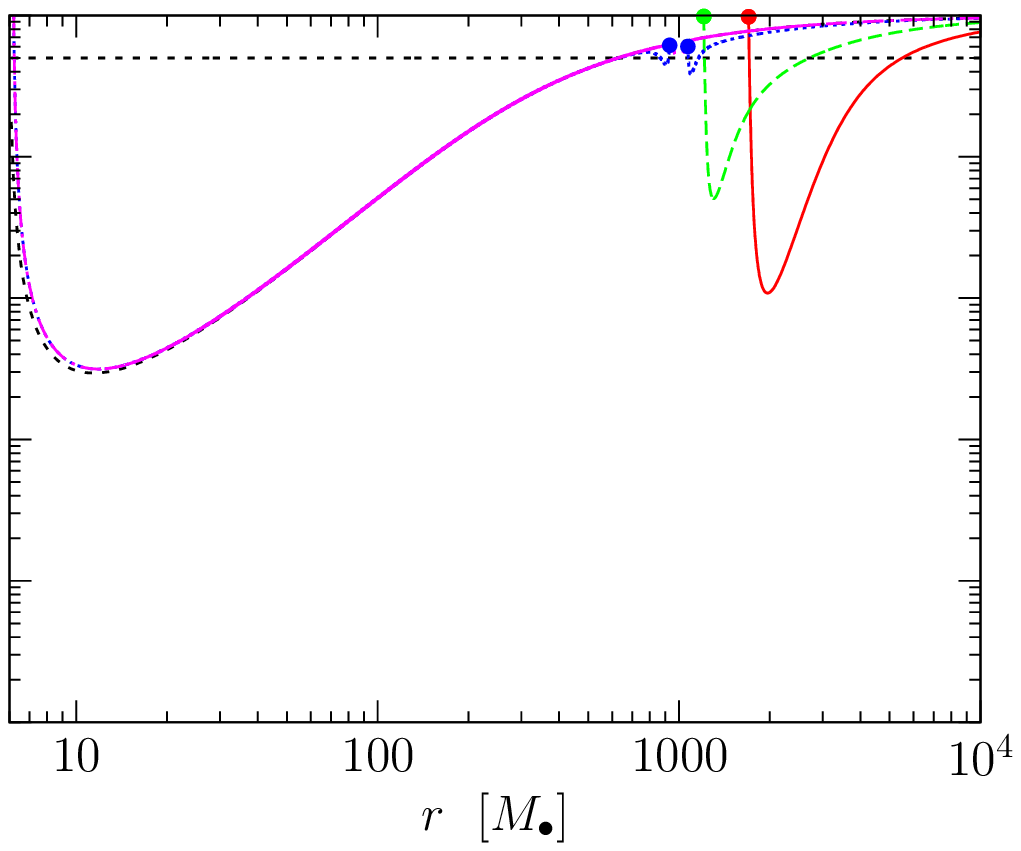}}
\caption{\label{f:disk}
Steady-state tidally heated disks with the secondary located at
$r_{\SCO}=\{35, 100, 1000 \}M_{\SMBH}$ (left, center, right panels)
for various mass ratios $q=m_{\SCO}/M_{\SMBH}=10^{-n}$ with
$n\in\left\{1,2,3,4\right\}$ (red, green, blue, magenta curves) for
$M_{\SMBH}=10^{7}\Msun$. The five different rows from top to bottom
show the local scaleheight, surface density, surface and midplane
temperature, surface brightness and tidal heating, and the gas to
total pressure ratio.  The magenta curve with $q=10^{-4}$ is similar
to a solitary disk without a secondary.  An inner cavity forms in the disk for $q=0.1$ and
$0.01$ for $r=1000\,M_{\SMBH}$ on the right panels.
The gas continuously overflows for the other cases shown in the figure, but
the surface density can still drop significantly near the
secondary's Hill sphere (marked by filled circles).
(For clarity, the curves are not connected
within the Hill sphere, where our model becomes invalid.)  }
\end{figure*}

In Figure~\ref{f:disk}, we present the most relevant physical
parameters in our steady-state disk as a function of radius (we
neglected $T_{\GW}$ for clarity in this figure) for several choices of
parameters.  For details on how these solutions were obtained in
practice, the reader is again referred to Paper I. As the figure
shows, the secondary acts as a hydro dam and causes gas to pile-up
(second row) and heat up (third row) relative to the unperturbed
solution. Interestingly, the disk brightens not only in the near zone
of the secondary, where the tidal effects dominate, but even much
farther away (see the panels in the 4th row showing the disk
brightness relative to the tidal heating rate). This is not
surprising, since viscosity needs to balance the tidal torque at the
boundary in equilibrium, and pushing the gas past this obstacle
requires a greatly accumulated local gas--mass, even far from this
region.  Our results are in qualitative agreement with the radial profiles
presented in \citet{2009MNRAS.398.1392L}.

Of the 12 cases shown in Figure~\ref{f:disk} with $r_{\SCO}=(35, 100,
1000) M_{\SMBH}$ and $q=(10^{-1},10^{-2},10^{-3},10^{-4})$, a gap
forms only in the two cases with the largest binary separations and
mass ratios: $r_{\SCO}=1000M_{\SMBH}$ with $q=0.1$ and $0.01$ (red and
green curves on the right panels).  In these cases, there is no inner
disk interior the secondary's orbit in the steady-state configuration
(if there was it would accrete onto the SMBH without being replenished
from the outside).  All other cases \emph{do not} have a
gap.\footnote{Here by ``gap'' we mean regions in which $\Sigma(r)=0$
  (i.e. $r<r_0$ where $r_0>r_{\SCO}+r_{\rm H}$ falls outside the Hill
  radius; see above).  Regardless, note that $\Sigma(r)$ can be
  greatly decreased in the overflowing solutions near the secondary.}
To avoid confusion, we note that the profiles are not shown within the
Hill sphere of the secondary in Fig.~\ref{f:disk}. Although our
working assumptions break down, the parameters are expected to
transition continuously across this region.

One may identify visually distinct zones in the radial profiles of the
physical parameters in Fig.~\ref{f:disk}.  At very large radii, which
we call the \emph{far zone}, the disk asymptotes to the unperturbed
state.  Interior to a certain radius, in the \emph{exterior middle
zone}, the scaleheight starts to deviate initially gradually then
quite rapidly as the radiation pressure becomes significant relative
to the gas pressure. Note that the transition to a radiation pressure
dominated disk (i.e. $\beta < 0.5$) is around $600 M_{\SMBH}$ without
a secondary.  However, the disk can become radiation pressure
dominated outside a secondary much farther out as shown by the 5th row
(c.f. dotted black line showing $\beta = 0.5$).  In the \emph{near
zone} of the satellite, the tidal effects become significant.  For
relatively massive binaries, the sharp knee in the tidal heating rate,
as seen in the 4th row, corresponds to the torque cutoff where the
midplane pressure gradient shifts the tidal effects out of resonance
($H\sim |r-r_{\SCO}|$).\footnote{Note that the changes seem
deceptively abrupt on a logarithmic scale.  To see this, consider
$y=(x-1)^2$, a smooth function, that exhibits a similar feature on a
log--log plot near $x=1$.}
After crossing the radius of the secondary's orbit, the accretion
velocity is more rapid than for an unperturbed disk by many orders of
magnitude (see Figure~\ref{f:disk-v} below).
The disk profile asymptotes to the unperturbed disk profile near the
inner boundary close to the SMBH.  Note however, that the tangential
velocity ($v_{t}\sim (r/M_{\SMBH})^{-1/2} \C$) and sound speed ($c_s
\sim (H/r) v_{t}$) are much larger than the radial velocity even in
this region.

We will discuss migration rates in detail in \S~\ref{s:migration}
below, but let us mention already which of the 12 cases shown in
Figure~\ref{f:disk} correspond to the well-known Type-I and II cases.
The two cases $r_{\SCO}=1000\,M_{\SMBH}$ with $q=0.1$ or $0.01$ have a
gap, and thus imply Type-II migration. The case with $q=10^{-4}$ and
$r_{\SCO}=35 M_{\SMBH}$ corresponds to Type-I migration, where the
disk structure lies close to the unperturbed case.  The 7 cases with
$q\geq 10^{-3}$ without a gap have a significantly perturbed
overflowing disk, which corresponds to the distinct new class of Type
1.5 migration. The remaining two cases with $q=10^{-4}$ and with
$r_{\SCO}=(100,1000) M_{\SMBH}$ are between Type I and 1.5. However,
comparing the disk-driven migration speed with the GW inspiral rate,
we find that GW emission is significant for many of these binaries, so
that the steady-state assumption is violated (see Fig.~\ref{f:gap}
below for the conditions under which this occurs).

\subsection{Analytic solutions}\label{s:analytical}

To generalize the numerical solutions given above for arbitrary
parameters, we derived in Paper~I approximate analytic formulas.
These solutions exist in different regions where either the tidal, the
viscous torques, or the angular momentum flux is negligible relative
to the other two terms in Eq.~(\ref{e:Tnu'}).  In particular, we
identify the two \emph{far zones}, well outside and far inside the
secondary's orbit, where the effects of the secondary are negligible,
a single \emph{middle zone} outside the secondary's orbit, where the
disk structure is greatly modified but where the tidal torque and
heating are negligible, and two \emph{near zones} just inside and
outside $r_{\SCO}$, where the tidal effects dominate.  We distinguish
two possible cases for the middle zone, depending on whether the disk
has a gap or if the disk is overflowing. In the external near zone, we
likewise have two possible behaviors, depending on whether the tidal
torque is unsaturated or saturated (i.e. whether the torque cutoff is
in play: $\Delta = |r-r_{\SCO}|$ or $H$ in Eq.~(\ref{e:Lambda}),
respectively).  We refer the reader to Paper~I for detailed
derivations of the analytic solution for the disk structure in each
zone; here we provide only the most important results, and discuss
their implications for SMBH binaries.

The physical parameters at radius $r$ in the disk are given by
\begin{equation}\label{e:X}
 X(r,r_{\SCO},\mathbf{p}) = C\, \alpha_{-1}^{c_1}\, \dot{m}_{-1}^{c_2}\, M_7^{c_3}\, r_{2}^{c_4}\, f_{-2}^{c_5} \, q_{-3}^{c_6} \, r_{\SCO 2}^{c_7} \,\Phi(r,r_{\SCO},\mathbf{p})
\end{equation}
where $X$ denotes any of $\{\Sigma, T_{c}, H, v_r, F, T_{\nu}\}$;
$r_{2}$ and $r_{\SCO 2}$ denote the radial distance from the primary
and the orbital radius of the secondary in units $100\, M_{\SMBH}$,
respectively; and $\Phi(r,r_{\SCO},\mathbf{p})$ denotes an extra
function of the parameters which is different in different zones.  The
constant parameters $C$ and $c_{i}$ and $\Phi$ are given explicitly in
Paper~I (see Table 1 therein).  Here we only discuss the structure of
the solution.

The steady-state viscous torque density $T_{\nu}$ in different regions
constitutes the backbone of the analytic solution.  Every physical
quantity in the disk follows directly from $T_{\nu}$ by simple
arithmetic relations.
To verify the analytic model with the numerical solutions of
\S~\ref{s:numerical}, it is sufficient to examine $T_{\nu}$ in the
various regions. We do this by generating the numerical solution for
randomly chosen constant model parameters
$(\alpha,\dot{m},f,M_{\SMBH},q,r_{\SCO})$, and compare these to the
analytic solution.  We find a good agreement between the two solutions
to within $20\%$ for a wide range of parameters for which $T_{\nu}$ is
substantially modified exterior to the $r_{\SCO}$.  We do not develop
analytic approximations in the opposite, weakly perturbed case,
because in that regime other physical processes which have been
neglected may be more significant (see Type-I migration in
\S~\ref{s:migration:type-I}).

The analytic solution is self-consistent in the strongly perturbed
case if $T_{\nu}$ matches continuously in the different zones outside
the Hill sphere.  In practice, note that $T_{\nu}$ increases in the
external near zone quite rapidly with radius, both in the tidal torque-saturated
and unsaturated regimes.  At larger radii, the tidal effects vanish
(middle zone) and $T_{\nu}$ is approximately constant.  If $T_{\nu}$
in the middle zone without a gap is lower than with a gap, then this
represents the correct overflowing solution; otherwise a gap forms.
Since for a steady-state disk,
the migration rate is simply proportional to $T_{\nu}$ in the middle zone
(see Eq.~30 in Paper~I),
this definition is equivalent to that based on the velocities (Eq.~\ref{e:vr0}).
If so, the secondary and the gas propagate inward with similar
velocities to maintain this configuration, while in the former case,
the secondary moves slower than the inflow speed of the gas. We
elaborate on these issues related to gap opening and migration in
\S~\ref{s:gap} and \ref{s:migration} below.

We emphasize that the disk parameters can differ dramatically from the
unperturbed (or far zone) values not only in the near zone, where the
tidal effects dominate, but also in the middle zone, where tidal
effects are already negligible. This is analogous to how the water
level is raised in a dam not just in the immediate vicinity of the dam
wall but even far away from the boundary, regions in which the local
hydrodynamics is explicitly independent of the wall.  The tidal
effects of the secondary are also short-range, but the corresponding
effect is communicated to distant regions by setting an effective
boundary condition. The middle region subtends a large radial range,
and is representative of the strongly perturbed disk.
Here, the viscous torque
$T_{\nu}^{\rm m}$ (in all three cases: with a gap, or overflowing with
saturated or unsaturated torques; see the corresponding solutions
$T_{\nu}^{\rm mg}$, $T_{\nu}^{\rm mos}$, and $T_{\nu}^{\rm mou}$ in
Table 1 of Paper~I) is approximately constant with radius as in
a \textit{decretion disk} \citep{1991MNRAS.248..754P}. It is
remarkable that the perturbation to the disk can be represented by a single number in the
middle zone. Here, the radial profile of the disk is independent of the details of the
tidal torque up to this constant factor.

\begin{figure*}
\centering
\mbox{\includegraphics{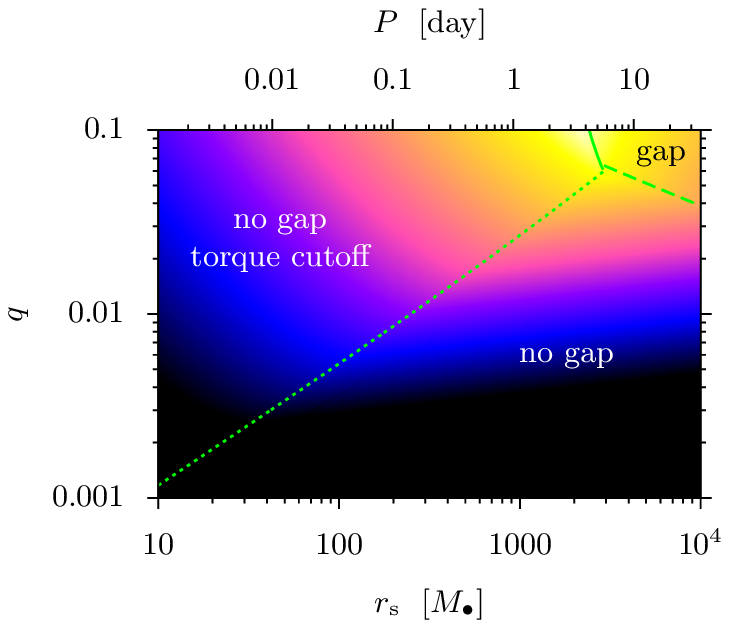}}
\quad
\mbox{\includegraphics{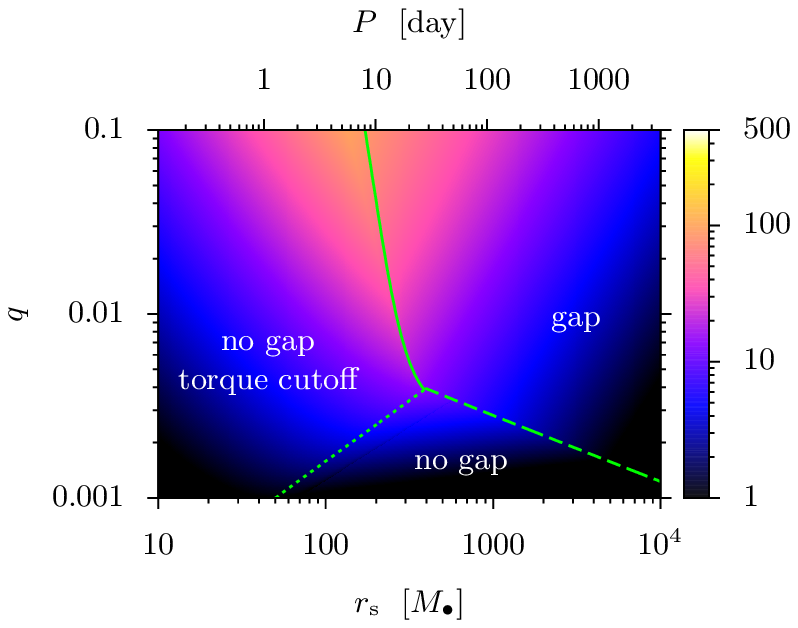}}
\quad
\mbox{\includegraphics{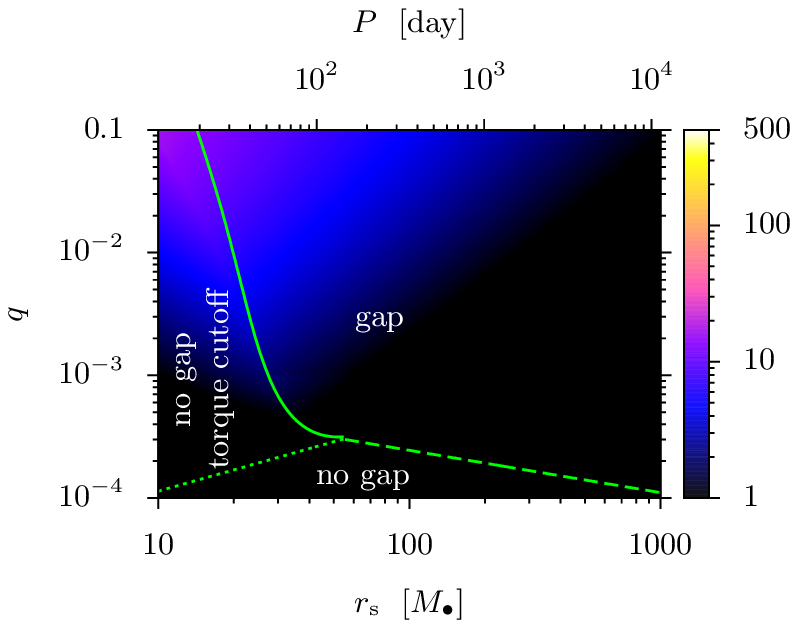}}
\caption{\label{f:k}
  The phase diagram of the disk for $M_{\SMBH}=10^5\Msun$ (top left)
  $10^7\Msun$ (top right) and $10^9\Msun$ (bottom) and for different
  binary semimajor axes, orbital periods, and mass ratios.  Colors
  show the relative brightening of the disk outside the secondary
  relative to the same region of a solitary disk. Three zones can be
  distinguished as marked: ``gap'' in which the tidal torque can
  support a gap against viscosity even in steady-state, ``no gap'' in
  which the gap overflows in steady-state with the tidal torque in the
  linear regime all the way to the Hill sphere or ``no gap with torque
  cutoff'' in which the torques are saturated outside the Hill sphere.
  We do not show comparable mass ratios ($q>0.1$) where the adopted
  perturbative tidal torque formula and the axisymmetric steady-state
  approximations are strongly violated.}
\end{figure*}

Rather than using $T_{\nu}^{\rm m}$, we introduce a more physically
revealing quantity, $k\equiv T_{\nu}^{\rm m}/T_{\nu 0}$, representing
the fractional change of the viscous torque relative to its
unperturbed value ($T_{\nu 0}=\dot{M}r^2\Omega$) in
the middle zone. This parameter, which we refer to as the {dimensionless angular momemntum flux or}
\emph{brightening factor},  sets the relative brightening of the disk and the increase in the scaleheight in the middle zone,
as well as the migration rate. It can take one of three values, depending on whether a gap
is open, or whether the gas overflows in the saturated torque
cutoff regime or in the unsaturated regime\footnote{The brightening factor
has a simple radial dependence in the middle zone $k\propto r^{-1/2}$.
We extrapolate the brightening factor in the middle zone to the location of the secondary, typically a
conservative estimate of the true brightening of the disk outside the secondary's orbit
as this neglects the excess brightening due to tidal heating in the near zone (see Paper I).}:
\begin{align}\label{e:kmg}
 k^{\rm mg}_{\SCO} &= 23\, \alpha_{-1}^{1/2} \dot{m}_{0.1}^{-3/8} M_7^{-3/4} q_{-3}^{5/8} \lambda^{-11/16} r_{\SCO 2}^{-7/8}\\\label{e:kmos}
 k^{\rm mos}_{\SCO} &= 0.97\, \alpha_{-1}^{-2/11} \dot{m}_{-1}^{-1} M_7^{1/22} f_{-2}^{5/22} q_{-3}^{5/11} r_{\SCO 2}^{39/44}
\nonumber\\ &\quad\times [-\W(-a)]^{13/11}\,,\\\label{e:kmou}
 k^{\rm mou}_{\SCO} &= 1.3\, \alpha_{-1}^{-2} \dot{m}_{-1}^{-1} M_7^{1/2} f_{-2}^{5/2} q_{-3}^{5/2} r_{\SCO 2}^{-1/4}
\nonumber\\ &\quad\times \left[ 1+\left( \frac{\delta r_{\rm i}}{r_{\SCO}} \right)^{1/3} \right]^{-115/24} \left(\frac{\delta r_{\rm i}}{r_{\rm H}}\right)^{-15/2}\,.
\end{align}
Here $\W(-a)$ is the Lambert W-function defined for $a>1/e=0.368$ approximately as
\begin{equation}\label{e:W}
 |\W(-a)|\approx -\ln(-a)+\ln(-\ln(a))
\end{equation}
where
\begin{equation}\label{e:adef0}
 a = 0.465\, \alpha_{-1} f_{-2}^{-5/4} q_{-3}^{-13/12} M_7^{-1/4} r_{\SCO 2}^{5/8} \left(\frac{\delta r_{\rm i}}{r_{\rm H}}\right)^{17/4}\,.
\end{equation}
In practice, $1\lesssim |\W(-a)|\lesssim 10$ for a wide range of parameters.
In the above equations, $\delta r_{\rm i}$ is the radial distance from $r_{\SCO}$ at which the
torque model breaks down near the secondary, for which we assume $\delta r_{\rm i}\sim r_{\rm
  H}$. Further,
$\lambda r_{\SCO}$ is the characteristic radius in the near zone outside the gap
where most of the tidal torque is exerted. In practice, $\lambda$ is of order unity given by Eq.~(\ref{e:lambdadef}) below.

The smallest of the three, $k_{\SCO}=\min( k^{\rm mg}_{\SCO}, k^{\rm mos}_{\SCO},
k^{\rm mou}_{\SCO})$, sets the state of the disk in the middle zone if
greater than one.\footnote{We do not consider the cases where either
  $k^{\rm mg}_{\SCO}$, $k^{\rm mos}_{\SCO}$, or $k^{\rm mou}_{\SCO}$ is less
  than one. In this case, the analytic solutions need further
  modifications (see Paper~I).}  Note that $k^{\rm mg}_{\SCO}$ increases
quickly with decreasing binary separation, $r_{\SCO}$, but interior to
some radius the gap closes and the {dimensionless angular momentum flux or brightening factor} is limited by the
near-zone torque cutoff, $k^{\rm mos}_{\SCO}$.  The brightening factor is
largest at the gap-closing boundary for comparable mass ratios.  The
torque cutoff does not impose a limitation if $q$ is sufficiently
small, but the gap can still close by shrinking to within the Hill
radius (in this case $k^{\rm mou}_{\SCO}$ sets the brightening factor).

Figure~\ref{f:k} shows the phase diagram of the disk for different
$r_{\SCO}$ and $q$, set by Eqs.~(\ref{e:kmg}--\ref{e:kmou}) for
$M_{\SMBH}=10^5$, $10^7$, and $10^9\Msun$.  Colors indicate the
brightening factor, $k_{\SCO}$.  The figure shows large deviations from a
solitary disk in many cases. The disk brightens in the middle zone by
a factor of up to $500$ or $100$ for a binary with $M_{\SMBH}=10^5$ or
$10^7\Msun$, respectively, which we discuss in
\S~\ref{s:observations}.  The brightening is less significant for
binaries with $10^{9}\Msun$. Note that GW emission is neglected here;
this modifies the picture at small radii significantly, especially for
$M_{\SMBH}=10^{9}\Msun$ (see Fig.~\ref{f:gap} below).

{We note that these solutions assume a steady-state. However,
in Paper~I we have shown that the steady-state assumption is strongly violated
by the inward migration of the secondary
in the case of a truncated disk with a central cavity if $\dot{M}$ is fixed
at the outer boundary \citep[see however discussion in \S~\ref{s:migration:type-II}]{1995MNRAS.277..758S}.
In contrast, steady--state is possible in the overflowing case
because there $k_{\SCO}$ decreases as the secondary moves inwards.
A global steady-state is possible if $k_{\SCO}\lesssim k_{\rm max}$ where
\begin{equation}\label{e:kmax}
 k_{\SCO \max} = 3.2\, |\gamma_{\Sigma {\SCO}}|^{5/19} \alpha_{-1}^{4/19} \dot{m}_{-1}^{-3/19} M_7^{6/19} q_{-3}^{5/19} r_{\SCO 2}^{-7/19}\,.
\end{equation}
A similar requirement for a global steady-state in a
radiation pressure dominated $\alpha$--disk is less restrictive.
For larger $k_{\SCO}$, an approximate steady-state with a fixed $\dot{M}$ may still hold locally in the inner regions
of the middle zone, but in this case, the outer parts of the middle zone
within $r\sim k_{\SCO}^2 r_{\SCO}$
cannot respond to the inward migration of the secondary and
the steady-state disk model becomes inaccurate there (see further discussion in
\S~\ref{s:migration:type-II}).
}

We finally comment on the stability of the accretion disk in the
middle zone.  Although the surface density increases as gas
accumulates outside the secondary as $k^{3/5}$, the radiation pressure
grows even more rapidly as $p_{\rm rad}\propto k$ in that region (see Eqs.~40--41 in Paper~I).
The net effect is to increase the Toomre-$Q$ parameter as $Q\propto
c_s/\Sigma \propto p_{\rm rad}/\Sigma \propto k^{2/5}$.  Therefore,
the enhanced radiation pressure dominates over the increase in surface
density, and makes the disk more stable against {axisymmetric} fragmentation relative
to an unperturbed disk without a secondary
\citep{2012MNRAS.420..705T}. Although gravitationally stable,
the disk may become dynamically unstable against
global non-axisymmetric perturbations if the disk develops
a steep pressure gradient in the near zone around a high-mass secondary
\citep{1985MNRAS.213..799P,1986MNRAS.221..339G}.
A proper stability analysis, and exploring its implications, is left to future
work.

\section{Gap opening}\label{s:gap}

Next we elaborate on the gap opening and closing conditions in more
detail.  Gap opening is traditionally examined by comparing the tidal
and viscous torques in an initially unperturbed steady-state thin
disk.  However, after a sufficient amount of gas has built up outside
the gap, the enhanced viscous torque may close the gap.  We discuss
these points in turn.

Gap opening also affects the migration rate of the secondary and has
several potentially observable signatures which we discuss in
\S~\ref{s:migration} and \ref{s:observations}.

\subsection{Initial {gap-opening}}

The velocity of the gas and the secondary can be computed starting
from Eqs.~(\ref{e:Tnu'}) and (\ref{e:SCO}) [see Paper I for
details]:
\begin{align}\label{e:vr}
v_{r} &= -\frac{\partial_r T_{\nu}}{2\pi r \Sigma \partial_r (r^2 \Omega)} +\frac{\Lambda}{\partial_r (r^2\Omega)}
= -\frac{\dot{M}}{2\pi r \Sigma}\,, \\
v_{\SCO r} &= - \frac{2 }{ m_{\SCO} \Omega_{\SCO} r_{\SCO}} \left(\int_0^{\infty} \partial_r T_{\rm d}\; dr + T_{\GW}\right)\,.
\label{e:vSCOr}
\end{align}
where negative values represent an inward motion.

Consider a secondary placed in an initially unperturbed disk, and
examine the conditions for the local gas flow to be reversed.  This is
equivalent to finding the region $v_{r}(r_0) \geq 0$ in
Eq.~(\ref{e:vr}), which is where the tidal torques dominate over the
viscous torques, or where the tidal timescale is smaller than the
viscous timescale in an unperturbed disk.  Substituting $T_{\nu}$ from
Eq.~(\ref{e:Tnu}) and $\Lambda$ from Eq.~(\ref{e:Lambda}), and
assuming that $r=r_{\SCO}+\Delta$ where $\Delta\ll r_{\SCO}$, setting
$v_{r}=0$ in Eq.~(\ref{e:vr}), we find
\begin{equation}\label{e:Deltagap}
\frac{\Delta_0}{r_{\SCO}} = \left(\frac{f}{3  k_{\nu 0}} \frac{q^2 r_{\SCO}^2 \Omega_{\SCO}}{ \nu_0} \right)^{1/3}\,,
\end{equation}
where $\nu_0$ is the unperturbed viscosity near the secondary, and
$k_{\nu 0} = \D \ln T_{\nu 0} / \D \ln \Delta = 1/2$ is the radial
exponent of the unperturbed viscous torque near $r_{\SCO}$.  Comparing
Eq.~(\ref{e:Deltagap}) with gas pressure dominated locally isothermal
hydrodynamical numerical simulations with no tidal heating, gives a
prefactor in the range $3k_{\nu 0}/f \sim 40$--$50$
\citep{2006Icar..181..587C}, which implies $f=0.03$--0.04.

\begin{figure*}
\centering
\mbox{\includegraphics[width=8.5cm]{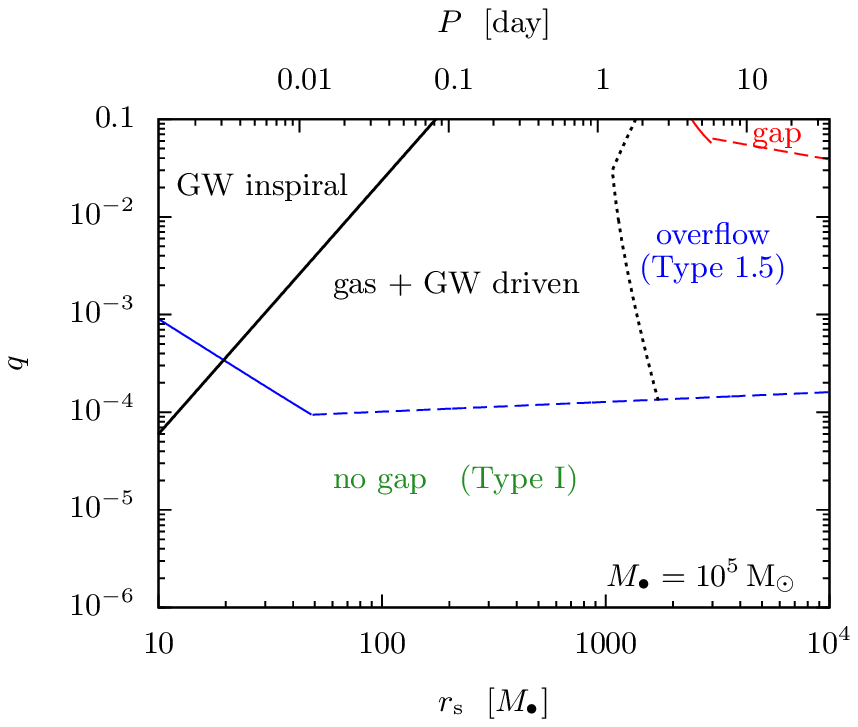}}
\quad\mbox{\includegraphics[width=8.5cm]{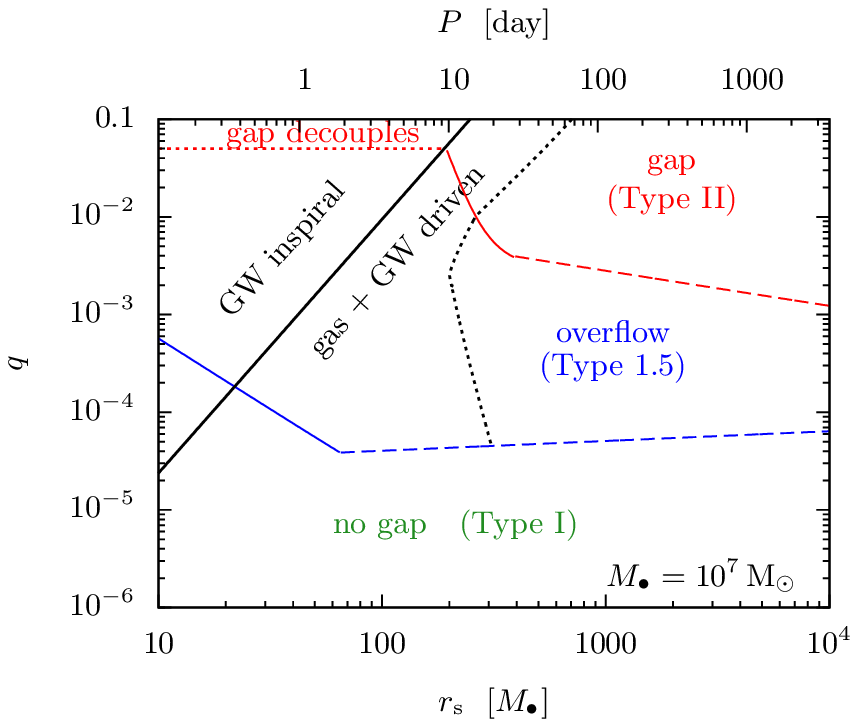}}\\
 \mbox{\includegraphics[width=8.5cm]{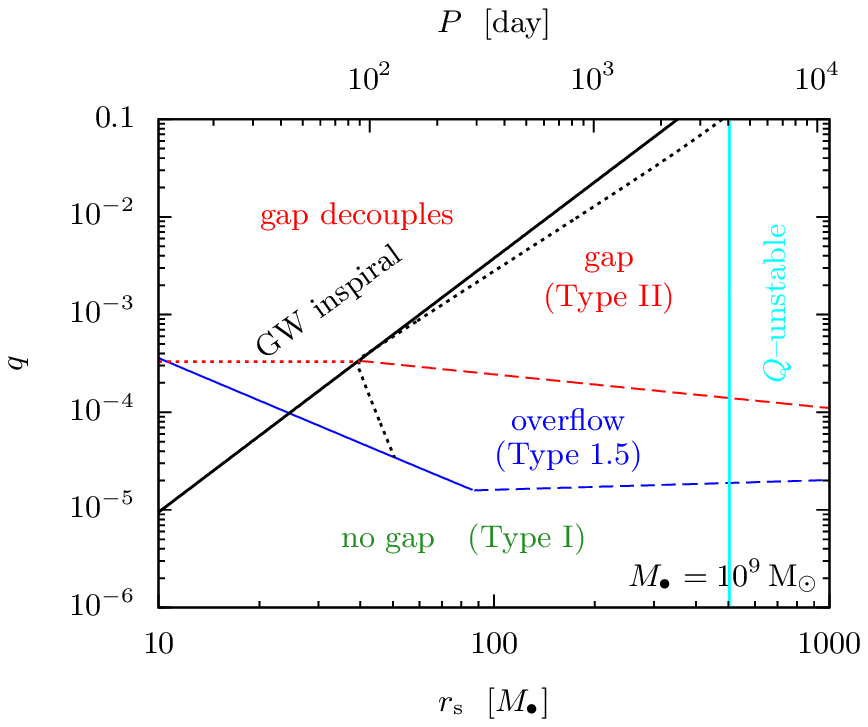}}
\caption{\label{f:gap}
  Ranges of binary separation $r_{\SCO}$ and mass ratio $q$ for which:
  (i) a gap can remain open in the disk; (ii) the gap refills after
  gas pileup (labeled as ``overflow''), and (iii) a gap cannot open. Different panels show different
 primary masses ($10^5$, $10^7$, and $10^9\Msun$) as labeled.
 The type of migration is shown
  in each region, including the region where GW losses start to become
  relevant (dotted line) and where they dominate (solid black line).
  Solid and dashed colored lines show the boundary where the tidal
  torque becomes saturated outside the Hill sphere (viscous filling,
  $\Delta=H$) and where the gas enters the Hill sphere without
  becoming saturated (gravitational filling, $\Delta=r_{\rm H}$),
  respectively, based on standard criteria using unperturbed disks.
  The gap (if present) decouples in the GW inspiral regime as shown.
  The disk is gravitationally unstable at large radii as marked.
}
\end{figure*}

Two conditions are required\footnote{{The gap opening conditions are sometimes written as
$\Delta_0\gtrsim r_{\rm H} \gtrsim H_0$, where $r_{\rm H} \gtrsim H_0$ is the condition for the azimuthal
perturbations to become nonlinear \citep{1997Icar..126..261W,1996ApJS..105..181K}.
We shall not require $r_{\rm H} \gtrsim H_0$ here over the two
conditions in Eq.~(\ref{e:gap}).}}
{for the tidal torques to initially truncate the disk and create a gap} after
inserting an object into the disk
\citep{1995MNRAS.277..758S,1997Icar..126..261W,2006Icar..181..587C}:
\begin{equation}\label{e:gap}
 \Delta_0 \gtrsim H_0 {~~~\rm and~~~} \Delta_0 \gtrsim r_{\rm H}\,.
\end{equation}
First, the maximum tidal torque, which corresponds to a distance
$\Delta_0 \sim H_0$, needs to exceed the viscous torque to keep gas
from flowing in.  Inside this distance, the tidal torque saturates due
to the ``torque cutoff'' and cannot counteract the viscous torque
\citep{1980ApJ...241..425G} (see also discussion below
Eq.~[\ref{e:Lambda}]).  Thus, gap opening requires, $\Delta_0 \gtrsim H_0$.
Second, the derivation of the torque formula assumed that the effect
of the secondary is a small perturbation to the gravitational
potential; this assumption breaks down interior to the Hill
sphere. Thus, consistency also requires $\Delta_0 \gtrsim r_{\rm H}$.  Gas
may be expected to accrete onto the secondary or to cross the gap, if
this condition is violated.  In addition to violating the conditions
in Eq.~(\ref{e:gap}), a gap may also close by 3D overflow if
$H \gtrsim r_{\SCO}$.  However, Ward
\citep{1986Icar...67..164W,1988Icar...73..330W} has shown that using
the 2D midplane torque gives a result typically within $20\%$ to the
vertically averaged torque \citep[see however][]{2005ApJ...619.1123J},
so that the gap closes by midplane inflow in typical cases, and
additional criteria are not necessary.

These gap opening criteria translate into bounds on the mass ratio
for fixed $r_{\SCO}$ \citep{1986ApJ...309..846L,2006Icar..181..587C} where
\begin{align}\label{e:qs0}
q \gtrsim q_{\rm s0} &= \sqrt{\frac{3 k_{\nu 0}}{f} \frac{\nu_0}{r_{\SCO}^2 \Omega_{\SCO}}}\left(\frac{H_0}{r_{\SCO}}\right)^{3/2}\,,~~\rm and\\
q \gtrsim q_{\rm u0} &= \frac{k_{\nu 0}}{f} \frac{\nu_0}{r_{\SCO}^2 \Omega_{\SCO}}\,,
\label{e:qu0}
\end{align}
If $q\lesssim q_{\rm s 0}$ the gap is closed by viscosity as the tidal
torque saturates due to the pressure gradients, and if $q\lesssim q_{\rm
  u0}$ it is closed by the near-field gravity of the secondary.  Thus,
after releasing an object in a radiation pressure dominated $\alpha$
or $\beta$ disk, a gap opens if \citep{2011PhRvD..84b4032K}
\begin{align}\label{e:q_alpha0}
 q_{\alpha0} \gtrsim & \max\left\{
4.5\times 10^{-5} \frac{\alpha_{-1}^{1/2}}{f_{-2}^{1/2}} \frac{\dot{m}_{-1}^{5/2}}{r_{\SCO 2}^{5/2}},
5.7\times 10^{-4} \frac{\alpha_{-1}}{f_{-2}} \frac{\dot{m}_{-1}^{2}}{r_{\SCO 2}^{2}}
 \right\}\\
 q_{\beta0} \gtrsim & \max\left\{
 2.0\times 10^{-5} \alpha_{-1}^{2/5} \dot{m}_{-1}^{17/10} M_7^{-1/10} f_{-2}^{-1/2} r_{\SCO 2}^{-29/20}\,,\right.\nonumber\\
&\quad\left.
 4.0\times 10^{-5} \alpha_{-1}^{4/5} \dot{m}_{-1}^{2/5} M_7^{-1/5} f_{-2}^{-1} r_{\SCO 2}^{1/10} \right\}
\label{e:q_beta0}
\end{align}
while for gas pressure dominated accretion disks
\begin{align}\label{e:q_gas0}
 q_{\rm gas0} \gtrsim & \max\left\{
 1.7\times 10^{-6} \alpha_{-1}^{1/4} \dot{m}_{-1}^{1/2} M_7^{-1/4} f_{-2}^{-1/2} r_{\SCO 2}^{1/8}\,,\right.\nonumber\\
&\quad\left.
 4.0\times 10^{-5} \alpha_{-1}^{4/5} \dot{m}_{-1}^{2/5} M_7^{-1/5} f_{-2}^{-1} r_{\SCO 2}^{1/10} \right\}\,.
\end{align}
Here, the first and second terms in each parentheses correspond to
$q_{\rm s0}$ {(Eq.~\ref{e:qs0})} and $q_{\rm u0}$ {(Eq.~\ref{e:qu0})}, respectively. Note that latter
is the same for gas and radiation pressure dominated
regimes for a $\beta$ disk {because this condition, Eq.~(\ref{e:qu0}),
is independent of $H$, and because $\nu$ is insensitive to radiation pressure in the $\beta$ model}.

Figure~\ref{f:gap}  shows the gap opening criteria for
$\beta$ disks with our standard parameters as a function of mass ratio
and binary separation for $M_{\SMBH}=10^{5,7,9}\Msun$.  The blue solid
and the dashed curves represent $q_{\rm s0}$ and $q_{\rm u0}$, respectively.
For smaller mass ratios, the disk is practically unperturbed.

\subsection{{Steady-state gap-closing}}
In the regime where a gap never opens (i.e. weakly perturbed disks, $q<q_{\rm crit 0}$), our solutions do not change the picture.
However, in the regime when a gap initially opens, the accumulation of gas outside the binary
changes $\Sigma$, $H$, $\nu$, and $k_{\nu}$ in the gap-opening conditions, Eqs.~(\ref{e:Deltagap}--\ref{e:gap}).
Let us now examine whether the gap can stay open on longer timescales once the disk and binary have
reached steady-state.

{
As the binary migrates inwards the local pressure and viscosity change even in an unperturbed disk, 
which may lead to gap opening for mass ratios of order $q\sim 10^{-4}$ within separations $r_{\SCO}\sim 10^3 M_{\SMBH}$
(see blue lines in Figure~\ref{f:gap}). 
However, the usual gap opening conditions Eqs.~(\ref{e:qs0}--\ref{e:qu0}) \citep{1986ApJ...309..846L,1995MNRAS.277..758S}
may fail in the long run and lead to gap-closing for a much wider range of masses and radii, 
if either the scaleheight, the viscosity,
or the steepness of the radial profile increases during the pile-up.
Note that  if only $H$ increased
by a factor $x$ while $\nu$ and $k_{\nu}$ were kept constant,
then the mass ratio for gap opening would increase by $x^{3/2}$
relative to the first terms in Eqs.~(\ref{e:q_alpha0}--\ref{e:q_gas0}).
Furthermore, if $\nu$ or $k_{\nu}$ changed by a factor $y$,
then the the critical mass ratio for gap opening would further increase
by an extra factor of $\sqrt{y}$ and $y$ in the first and second terms
in Eqs.~(\ref{e:q_alpha0}--\ref{e:q_gas0}), respectively.
Thus, the combination of these effects shifts the solid and dashed blue lines
in Figure~\ref{f:gap} upwards by a factor $x^{3/2}y^{1/2}$ and $y$, respectively.
We determine
the actual value of $x$ and $y$ in the self-consistent steady-state model to derive
the necessary conditions for gap opening next.}

{In a self-consistent decription, }gas accumulation outside the gap leads to the following effects.
First, the gravitational torque of a denser disk drives the migration of the secondary faster,
increasing $|v_{\SCO r}|$ (see Eqs.~\ref{e:Td} and \ref{e:vSCOr}). For a fixed accretion rate, $\dot{M}=2\pi r |v_{r}| \Sigma$, the increase
of $\Sigma$ leads to the decrease of $|v_{r}|$, analogous to the slowdown of the flow velocity in a river upstream a dam.
The pressure and kinematic viscosity increases due to the pile-up, and leads to a smaller gap.
The tidal heating in the gas outside the disk increases the pressure and scaleheight further which can quench the tidal torque
(see torque cutoff in Eq.~\ref{e:Lambda}).
Ultimately, the combination of these effects can have several outcomes.
\begin{enumerate}
\item If the mass of the secondary is very small, the disk is weakly
  perturbed (no gap is present in particular) and the object exhibits
  Type-I migration as usual (see \S~\ref{s:migration}).
\item The other extreme limit is Type-II migration, where $|v_{\SCO
    r}| = |v_{r}(\lambda r_{\SCO})|/\lambda$ is reached so that the disk and the
  secondary spiral inward in a self-similar way (see Eq.~\ref{e:vr0}).
  Here the relative gap size, $\lambda=r_{\rm g}/r_{\SCO}$, is
  somewhat decreased after the gas has accumulated outside the gap,
  but it is still much larger than the Hill radius.
\item An intermediate possibility is that the gap size shrinks over
  time to within the secondary's Hill sphere as gas accumulates
  outside the gap, and eventually closes.
  The equilibrium steady-state configuration in this continuously
  overflowing case has an increased density and pressure exterior to
  the secondary's orbit (see disk profiles in Fig.~\ref{f:disk} for
  $q\gtrsim 10^{-3}$).
\item Finally, it is also possible that the disk becomes geometrically
  so thick ($H>r$) and/or luminous ($L>L_{\rm Edd}$) to drive a wind
  or a three dimensional inflow across the orbit of the secondary.
\end{enumerate}
{A necessary condition for the tidal torques to sustain a cavity in the disk} is to satisfy
$|v_{\SCO r}| = |v_{r}(\lambda r_{\SCO})|/\lambda$, with the gap closing if and when
$|v_{\SCO r}|$ falls below this value. Here $v_{\SCO r}$, $v_{r}(\lambda r_{\SCO})$, and
$\lambda$ represent the final steady-state equilibrium values.
This condition can be expressed directly with the viscosity in the
disk as in Eqs.~(\ref{e:qs0}--\ref{e:qu0}), but with the
viscosity and scaleheight significantly different from their original
values. We utilize the analytic solution to the complete disk model
(\S~\ref{s:analytical}) as a function of the parameters $(\alpha,
\dot{m}, M, f, q, r_{\SCO})$ and calculate whether these conditions
are satisfied. In practice, we calculate the corresponding range of
$q$ and $r_{\SCO}$ by requiring that the viscous torque at the outer
boundary of the near zone be equal to that at the inner edge of the
middle zone with a gap assuming unsaturated and saturated tidal
torques, respectively (see Paper~I for details).  This is equivalent
to finding {$q$ or $r_{\SCO}$ for gap opening at which the dimensionless
angular momentum flux of an overflowing solution becomes first equal to that
of the model with a gap. At the gap opening/closing transition} $k^{\rm mg}=\min\{
k^{\rm mos}, k^{\rm mos}\}$ in Eqs.~(\ref{e:kmg}--\ref{e:kmou}).
Assuming radiation pressure
dominated $\beta$ disks and negligible GW losses, we find that
{a truncated disk with an inner cavity forms if}
\begin{align}\label{e:gap-qu}
 q_{\rm u} &\gtrsim 4.7\times 10^{-3}\,\alpha_{-1}^{4/3} M_{7}^{-2/3} f_{-2}^{-4/3} \dot{m}_{-1}^{1/3} r_{\SCO 2}^{-1/3}\,,~~{\rm and}\\
 r_{\rm s s} &\gtrsim 600\,M_{\SMBH}\, \alpha_{-1}^{12/31} f_{-2}^{-4/31} q_{-3}^{3/31} M_{7}^{-14/31} |\W(-a)|^{-104/155}
\label{e:gap-rs}
\end{align}
for unsaturated and saturated tidal torques, respectively (see
Eqs.~\ref{e:W}--\ref{e:adef0} for $\W$ and $a$).  Here the first and
second conditions are analogous to $\Delta \gtrsim r_{\rm H}$ {(or Eq.~\ref{e:qu0})} and $\Delta \gtrsim
H$ {(or Eq.~\ref{e:qs0})}, respectively, but account for the
self-consistent steady-state radial variations in $H$, $\Sigma$, and
$\nu$ when calculating the viscous and integrated tidal torques {for a $\beta$ disk}.
Note that Eq.~(\ref{e:gap-rs}) is
defined only for $a>1/e=0.368$, otherwise the tidal torque is in the
unsaturated state.  Since $a$ depends on $q_{\SCO}$ and $r_{\SCO}$,
Eq.~(\ref{e:gap-rs}) is a nonlinear equation for $r_{\SCO}$ for a
fixed $q_{\SCO}$. However, the dependence is logarithmically weak,
typically the last factor in Eq.~(\ref{e:gap-rs}) is between 0.3 and
0.8, making the saturated gap-closing radius vary between
$(200$--$400) M_7\,M_{\SMBH}$ for all $q$ for the fiducial disk
parameters.

As mentioned above, for our {overflowing} solutions to be physically
self-consistent, we must also require $H\lesssim r$ and $L \lesssim L_{\rm Edd}$, {otherwise
the radiation pressure would drive an outflow}. In
\S~\ref{s:observations}, we show that these two conditions are
approximately equivalent.  Assuming there is a gap, a radiation pressure dominated
$\beta$ disk becomes thick ($H>r$) for mass ratios
\begin{equation}
 q_{\rm thick} \gtrsim 5.4\times 10^{-3}\, \alpha_{-1}^{-4/5} \dot{m}_{-1}^{-1} M_7^{6/5} \lambda^{7/2} r_{\SCO 2}^{3}\,.
\end{equation}
We find this condition to be less restrictive than $|v_{\SCO r}| =
|v_{r}|/\lambda$ for unequal masses with $q<0.1$.  The disk does not
become geometrically thick if a gap is opened, for our standard disk
parameters for $q<0.1$.  The disk thickness is even smaller for radiation
pressure dominated $\alpha$ disks {implying that overflow occurs before
the disk drives a wind in this case too. These conclusions can be understood
by setting $H_0/r_s\sim 1$ and $\nu_0 \sim \alpha c_s H \beta^b \sim \alpha H^2 \Omega_{\SCO} \beta^b
\sim \alpha r_{\SCO}^2 \Omega_{\SCO} \beta^b$
in the simple gap opening conditions, Eq.~(\ref{e:qs0}--\ref{e:qu0}), which
implies $q_{\rm thick} \gtrsim (3k_{\nu 0}/f)^{1/2} \alpha^{1/2} \beta^{b/2} $.
Since the constant of proportionality $3k_{\nu 0}/f$ is of order $40-50$ in the nonlinear regime
\citep{1986ApJ...309..846L,2006Icar..181..587C}, there is no secondary mass that satisfies
this condition for $\alpha>0.025$ for an $\alpha$ disk with $b=0$.
We note that \citet{2012arXiv1205.5017R}
arrived at the opposite conclusion using the same argument but setting $3k_{\nu 0}/f\sim 1$,
claiming that the disk typically becomes thick and would drive an outflow
before the gap may close due to overflow.
Furthermore, note that the tidal dissipation effects
and the steepening of the angular momentum flux
profile $k_{\nu}>k_{\nu 0}$ help further to make the disk prone to overflow before the disk becomes
very thick.}

The red line in Fig.~\ref{f:gap}
shows the boundary where the gap remains open in
steady-state and where the disk overflows. Remarkably, the gas
overflows in steady-state and closes the gap for $q\gtrsim 10^{-4}$
for all {binary separations} below $1000\,M_{\SMBH}$, and gaps remain open only for
much larger secondary masses.

The gap-closing criteria derived above are independent of the initial
conditions of the disk assuming that the disk is
strongly perturbed and relaxes to steady-state. {However,
in Eq.~(\ref{e:kmax}) we have shown that this condition does not always
hold near the gap-closing boundary and a time-dependent study is
necessary for a more accurate description of gap closing. Furthermore, }
GW-emission becomes more and more efficient for decreasing binary
separations.  Figure~\ref{f:gap} shows regions where
GW emission starts to become relevant and where it dominates. In the
region marked ``gas+GW driven'', the GW-inspiral is slower than the
unperturbed viscous accretion speed but faster than the steady-state
gas overflow. We did not consider this regime in detail, but Paper~I
indicates that there may still be a build-up of gas mass, and possibly
overflow here, albeit at a slower rate than if GW emission was
neglected. In the region marked as ``GW inspiral'' in
Fig.~\ref{f:gap}, steady-state overflow is prevented
by the faster GW-inspiral.  In this region, the actual state of the
disk depends on the initial condition.  If a gap is open when the
binary crosses the critical radius where GW-emission becomes dominant,
then the gap decouples from the binary and remains empty until the
merger. Figure~\ref{f:gap} shows that this occurs if
the secondary mass is larger than approximately $m_{\SCO}\gtrsim
4\times 10^{5}\,\Msun$.  For these masses, an X-ray afterglow is
activated once the gas refills the gap on the viscous timescale,
typically many years after a GW event
\citep{2005ApJ...622L..93M,2010AJ....140..642T,2010PhRvD..82l3011L,2010ApJ...714..404T,2010PhRvD..81b4019S}.
However, for smaller secondary masses, Fig.~\ref{f:gap} shows that the
gap overflows and closes before entering the GW inspiral phase.  We
conclude that a gap may not be present, and significant X-ray emission
accompanies the GW-emitting stage for these {\it LISA/NGO} sources. The EM
counterpart may be modulated by the GW--driven binary
\citep{2002ApJ...567L...9A,2010MNRAS.407.2007C,2011PhRvD..84b4024F,2011CQGra..28i4020B,2012ApJ...744...45B,2012arXiv1204.1073N,2012MNRAS.tmpL.436B,2012arXiv1203.6108G,2012arXiv1207.3354F},
but future work is
necessary to explore this possibility in more detail with initial
conditions consistent with the overflowing solution.

\section{Migration}\label{s:migration}
\begin{figure*}
\centering
\mbox{\includegraphics[height=4.4cm]{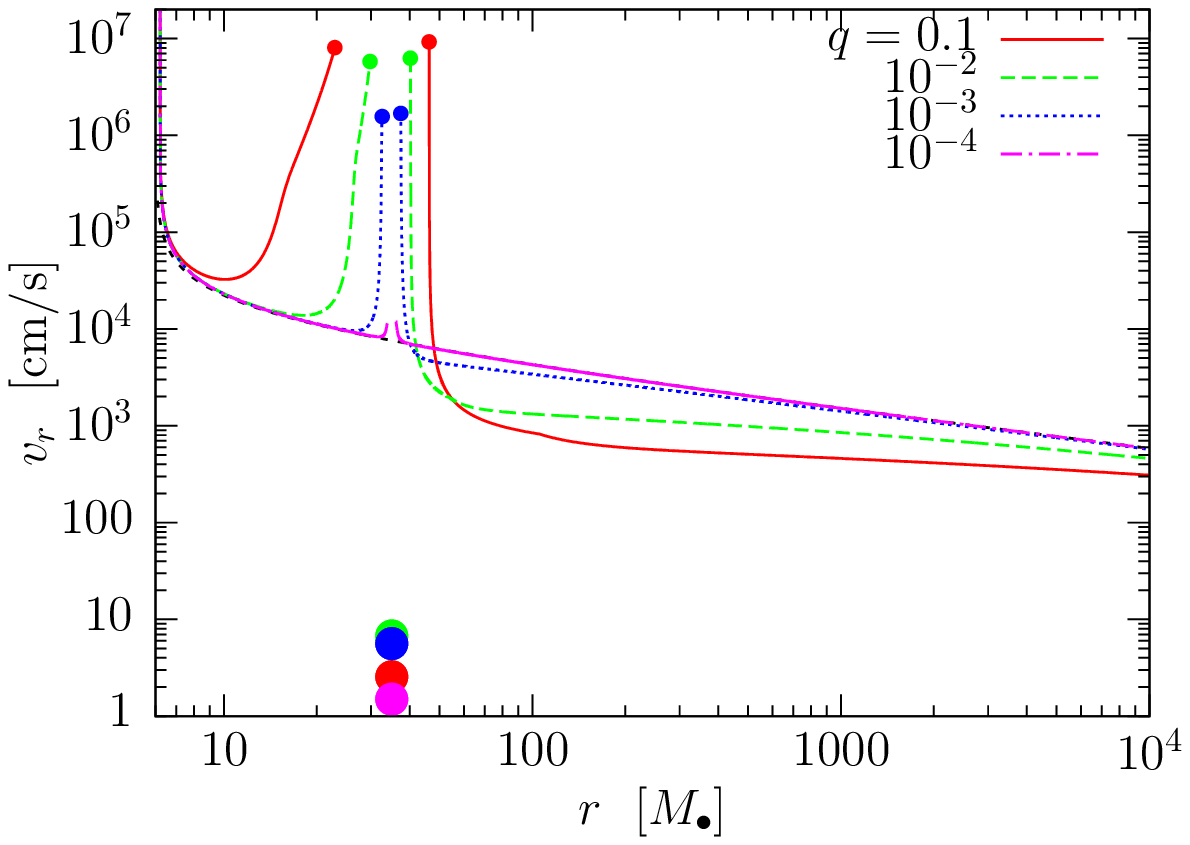}}
\mbox{\includegraphics[height=4.4cm]{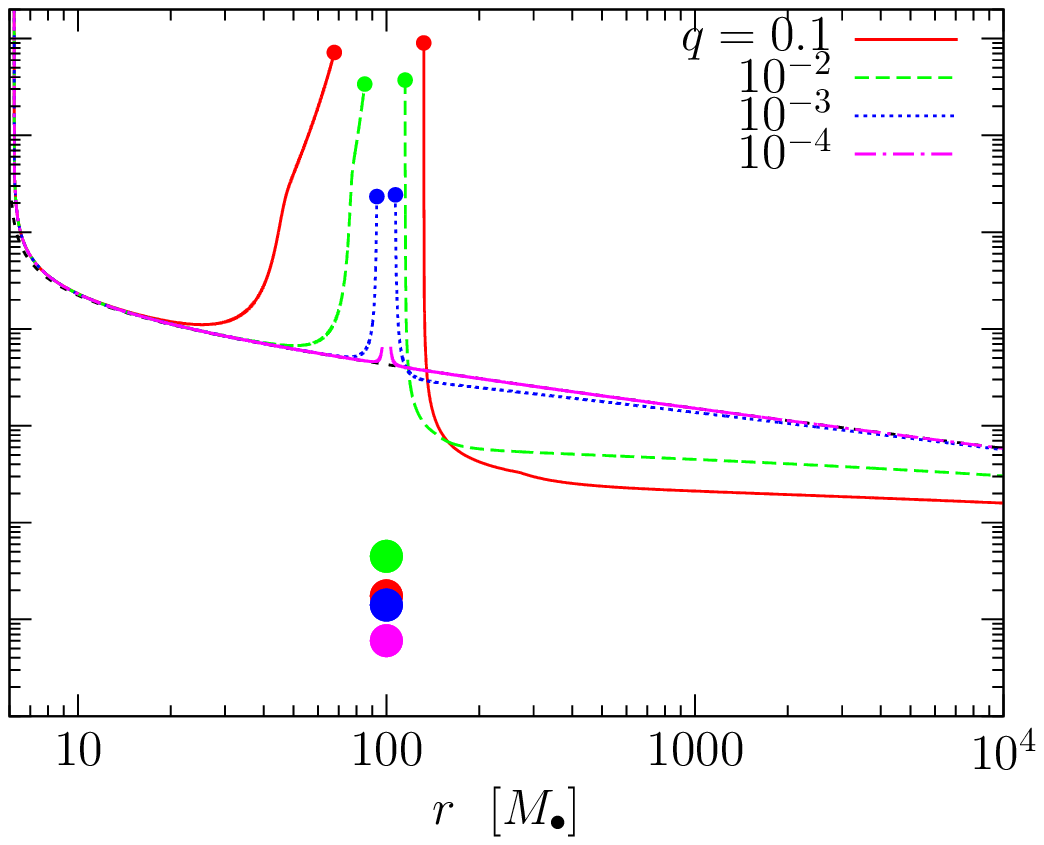}}
\mbox{\includegraphics[height=4.4cm]{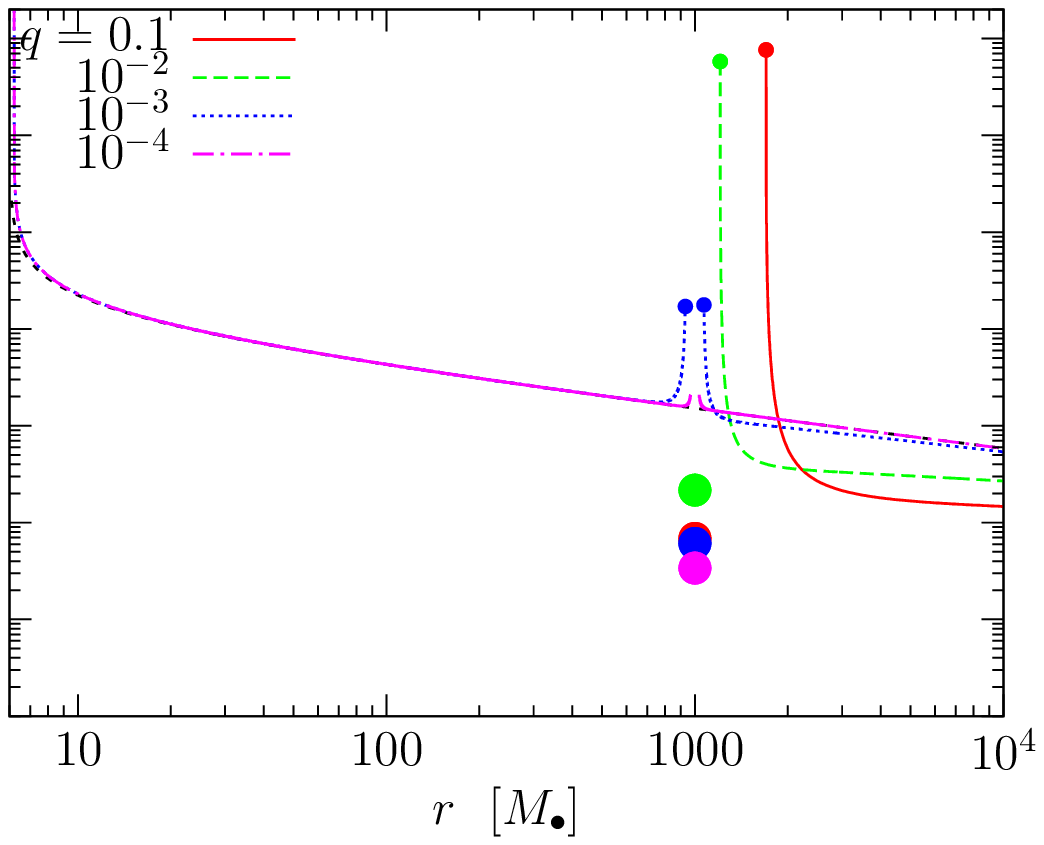}}
\caption{\label{f:disk-v}
  The accretion speed for the disks shown in Fig.~\ref{f:disk} and the
  migration speed of the secondary neglecting GW emission.  }
\end{figure*}

The migration rate of the secondary, $v_{\SCO r}$, follows from
Eq.~(\ref{e:vSCOr}), where $\partial_r T_{\rm d}$ is given in terms of
$\Sigma$ and $H$ using Eq.~(\ref{e:Td}--\ref{e:Lambda}) for which we
use the steady-state solution derived above. The result can be
expressed in terms of the dimensionless {angular momentum flux}
$k_{\SCO}$ in Eqs.~(\ref{e:kmg}--\ref{e:kmou}), as
\begin{equation}\label{e:vSCOr2}
v_{\SCO r} = -2 k_{\SCO} \frac{ \dot{M} r_{\SCO}}{m_{\SCO}} - \frac{64}{5} q \R_{\SCO}^{-3}\,,
\end{equation}
where the first and second terms correspond to disk-driven migration
and to GW-driven inspiral, respectively.

Figure~\ref{f:disk-v} shows the gas velocity as a function of radius,
and the migration rate of the secondary for the numerical solutions
corresponding to Fig.~\ref{f:disk}.  The disk has a gap for $q=0.1$
and 0.01 when $r_{\SCO}=10^3\,M_{\SMBH}$, where the nearby gas mass
and the secondary move radially at a similar speed.
{In this case, the secondary and the disk move inwards to maintain
a ratio $v_r(\lambda r_{\SCO})/v_{\SCO r}=\lambda$, where
$\lambda r_{\SCO}$ is the characteristic radius where the tidal torque is exerted.
The increase of the gas velocity near the gap edge is due to the fact that the
steady-state assumption is strongly violated there (see further discussion in
\S~\ref{s:migration:type-II}).
At smaller binary separations, the gas can flow across the
barrier represented by the secondary, the disk is no longer truncated, and
the migration rate is much slower than the gas velocity.}

The gas--secondary interaction has a different character when the mass
of the secondary is so small that it does not perturb the surface
density and scaleheight profile significantly (Type-I), and when it is
so large that a gap opens (Type-II). We first discuss these two
limiting cases below, and then turn to the intermediate state with a
steady-state overflowing disk {with a pile-up} (which we label ``Type-1.5'').

\subsection{Type-I migration}\label{s:migration:type-I}

Let us first consider a secondary with a mass so small that it makes
only a small change in the surface density and scaleheight profile
$\Sigma_0$ and $H_0$ {and the azimuthal density perturbations
are in the linear regime (i.e. $r_{\rm H} < H$).}
Here we estimate the corresponding Type-I
migration rate using the adopted simple tidal torque model,
Eq.~(\ref{e:Lambda}).\footnote{{
The tidal torque coefficient in Eq.~(\ref{e:Lambda}) is $f=0.80$ in this case, see \citet{1980ApJ...241..425G} and Paper~I.}}
For this estimate we extrapolate the torque to
within the Hill sphere.

These estimates are included for completeness here, but we emphasize
that they are very sensitive to the approximations used in the tidal
torque and the disk physics.  The true Type-I migration rate (even the
sign!) can be significantly different for many reasons including:
thermal effects \citep{2006A&A...459L..17P,2008A&A...485..877P}, heat
diffusion \citep{2011MNRAS.410..293P}, inclination
\citep{2011A&A...530A..41B}, turbulence and MHD effects
\citep{2004MNRAS.350..849N,2004ApJ...608..489L}, nonlinearities
\citep{2011ApJ...741...57D}, resonance overlaps
\citep{2012ApJ...747...24R}, and gas in horse-shoe orbits (Type-III
migration) \citep{2010MNRAS.401.1950P}.  3D effects for relatively
thin disks are less significant \citep{2002ApJ...565.1257T}.

We assume that near $r_{\SCO}$, $\Sigma =\Sigma_{\SCO}
(r/r_{\SCO})^{\gamma_{\Sigma}}$ and $H= H_{\SCO}
(r/r_{\SCO})^{\gamma_{H}}$, where the exponents ${\gamma_{\Sigma}}$
and $\gamma_{H}$ may be slightly different inside and outside the
secondary's orbit due to the tidal effects, $\gamma_{\Sigma, H}^{\rm
  ni}$ and $\gamma_{\Sigma, H}^{\rm ne}$, respectively.  For
unperturbed disks, $\gamma_{H}\approx 0$ and $\gamma_{\Sigma}\approx
3/2$ and $\gamma_{\Sigma}\approx -3/5$ for $\alpha$ and $\beta$ disks
respectively in the radiation pressure dominated case, while in the
gas pressure dominated case $\gamma_{H}\approx 21/20$ and
$\gamma_{\Sigma}\approx -3/5$ (see Table~1 in Paper~I and Eq. (5) in
\citealt{2009ApJ...700.1952H}).  Note that the scaleheight gradient
can be expressed in terms of the midplane temperature gradient
$T_c\propto r^{\gamma_T}$ as $\gamma_{H}=3 - \gamma_{\Sigma} +
4\gamma_{T}$ for $\beta\ll 1$ and $\gamma_{H} = \frac{3}{2} +
\frac{1}{2}\gamma_{T}$ for $\beta\sim 1$.

The tidal torque in Eq.~(\ref{e:vSCOr}) can be integrated analytically
in powers of the small quantity $H/r_{\SCO}$ in the four domains of
Eq.~(\ref{e:Lambda}), respectively, where $r>r_{\SCO}$ or $r<r_{\SCO}$
(i.e. interior vs. exterior zone) and $\Delta = H$ or $|r-r_{\SCO}|$
(i.e. saturated vs. unsaturated torque).  To first beyond leading
order,
\begin{align}
v_{\SCO r, \rm I,{\rm iu}} &= +\frac{1}{3} v_{{\rm I}, -3} - \left(2 + \frac{\gamma_{\Sigma}^{\rm ni}}{2} -\gamma_{H}^{\rm ni}\right) v_{{\rm I}, -2}
\,,\\
 v_{\SCO r, \rm I,{\rm is}} &= v_{{\rm I}, -3} - \left(2 + \frac{\gamma_{\Sigma}^{\rm ni}}{2} -\gamma_{H}^{\rm ni}\right) v_{{\rm I}, -2}
\,,\\
 v_{\SCO r, \rm I,{\rm es}} &=  -v_{{\rm I}, -3} - \left(\frac{\gamma_{\Sigma}^{\rm ne} }{2} -\gamma_{H}^{\rm ne}\right) v_{{\rm I}, -2}
\,,\\
 v_{\SCO r, \rm I,{\rm eu}} &= -\frac{1}{3} v_{{\rm I}, -3} - \left(\frac{\gamma_{\Sigma}^{\rm ne}}{2} -\gamma_{H}^{\rm ne}\right) v_{{\rm I}, -2}
\,,
\end{align}
where we have expressed the results in terms of
\begin{equation}\label{e:vn}
 v_{{\rm I}, n} = 2\pi f \frac{q}{M_{\SMBH}} \Sigma_{\SCO}  r_{\SCO}^3 \Omega_{\SCO} \left(\frac{H_{\SCO}}{r_{\SCO}}\right)^n\,.
\end{equation}
In particular for unperturbed $\beta$--disks,
\begin{equation}\label{e:v-TypeIa}
 v_{{\rm I}, -2} =
 \left\{
\begin{array}{l}
440 {\,\rm cm}{\,\rm s}^{-1} \alpha_{-1}^{-4/5} \dot{m}_{-1}^{-7/5} f_{-2} q_{-3} M_7^{6/5} r_{\SCO 2}^{29/10} \\\quad~~{\rm if}~\beta\ll 1\,,\\
1.2\times 10^4 {\,\rm cm}{\,\rm s}^{-1} \alpha_{-1}^{-3/5} \dot{m}_{-1}^{1/5} f_{-2} q_{-3} M_7^{7/5} r_{\SCO 2}^{4/5} \\\quad~~{\rm if}~\beta\sim 1\,.
\end{array}
\right.
\end{equation}
Thus, all four domains contribute to the migration rate in a nonnegligible way.
The total interior and exterior torques are
\begin{align}
v_{\SCO r, \rm I,{\rm i}} &= v_{\SCO r, \rm I,{\rm iu}} + v_{\SCO r, \rm I,{\rm is}} =
 \frac{4}{3} v_{{\rm I}, -3} - \left(4 + \gamma_{\Sigma}^{\rm ni} - 2\gamma_{H}^{\rm ni}\right) v_{{\rm I}, -2}
 \,,\\
   v_{\SCO r, \rm I,{\rm e}} &= v_{\SCO r, \rm I,{\rm es}} + v_{\SCO r, \rm I,{\rm eu}} =
 -\frac{4}{3} v_{{\rm I}, -3} - \left(\gamma_{\Sigma}^{\rm ne} -2 \gamma_{H}^{\rm ne}\right)  v_{{\rm I}, -2}
\,.
\end{align}

When combining the torques of the interior and exterior zones, the
leading order term, proportional to $v_{{\rm I}, -3}$, drops out,
irrespective of the exponents.  This cancellation makes Type-I
migration very sensitive to the local disk physics.  Denoting average
(over the inside and outside near zones) quantities by
$\bar{\gamma}_{\Sigma}$ and $\bar{\gamma}_{H}$,
\begin{align}\label{e:v-TypeIb}
 v_{\SCO r, \rm I} &= v_{\SCO r, \rm I,{\rm i}} +   v_{\SCO r, \rm I,{\rm e}} =
  -\left(4+ 2\bar{\gamma}_{\Sigma} - 4 \bar{\gamma}_{H}\right) v_{{\rm I}, -2}
\\
&= \left\{
\begin{array}{ll}
-7 v_{{\rm I}, -2} 
 &{\rm~if~~} \beta\ll 1,\; b=0\,,\\
-\frac{14}{5} v_{{\rm I}, -2}
&{\rm~if~~} \beta\ll 1,\; b=1\,,\\
\frac{7}{5}  v_{{\rm I}, -2} 
&{\rm~if~~} \beta\sim 1\,.
\end{array}
\right.
\end{align}
The last equality corresponds to the exponents in unperturbed $\alpha$
and $\beta$--disks as given above.

Equation~(\ref{e:v-TypeIb}) can be understood as follows. The
migration rate is a consequence of the opposing repulsive tidal
effects of the inner and the outer disks. Since the local disk mass is
proportional to $r^2 \Sigma \propto r^{2+\gamma_{\Sigma}}$, the
migration rate is directed inward for a constant thickness disk
($\gamma_{H}=0$), if and only if the local disk mass increases
outward. However, the scaleheight sets the scale at which the torque
is suppressed, and acts to reduce the effect.  For gas pressure
dominated disks, these estimates match the magnitude
of the Type-I migration rate of more accurate local torque and
hydrodynamical models of \citet{2002ApJ...565.1257T} and
\citet{2010MNRAS.401.1950P} if $f \sim 0.3$ in Eq.~(\ref{e:Lambda})
and (\ref{e:vn}).  A different overall offset, mostly due to
co-rotation torques and pressure gradient effects, directs the
migration inward in those models for a gas-pressure dominated
disk.  However, recent numerical simulations of
non-isothermal optically thick disks with tidal heating find outward
migration \citep{2006A&A...459L..17P}, and for MHD turbulent disks the
sign of the migration oscillates stochastically
\citep{2004MNRAS.350..849N}.  This highlights the extreme sensitivity
of Type-I migration to subtleties in the disk physics for low-mass
secondaries.

\subsection{Type-II migration}\label{s:migration:type-II}
\begin{figure*}
\centering
\mbox{\includegraphics{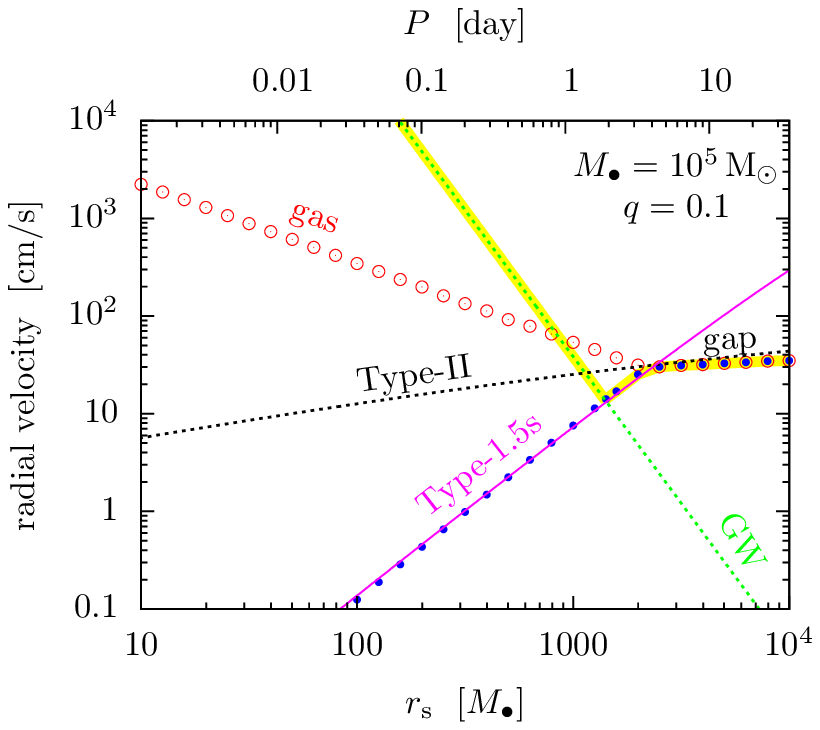}}\qquad
\mbox{\includegraphics{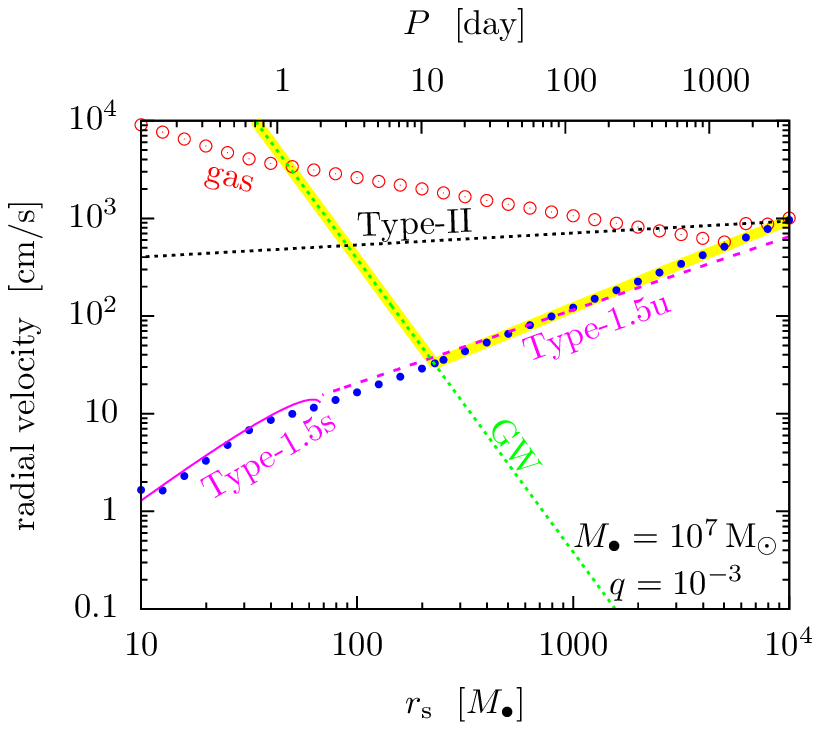}}
\caption{\label{f:migrationrate-r}
  Migration rate of the secondary (small blue bullets), radial
  velocity of the gas (red circles) from the numerical solution
  neglecting GW emission for $(M_{\SMBH},q)=(10^5\Msun, 0.1)$ (left)
  and $(10^7\Msun, 0.001)$ (right panel).  Analytic migration rates
  are also shown as labeled (Type-II, 1.5, GW) as a function of
  orbital radius.  The magenta line shows the saturated and
  unsaturated Type-1.5 migration rates, black dotted line is the
  Type-II rate.  A gap is present when the blue bullets and red
  circles overlap.  The green dotted line shows the GW inspiral rate.
  The numerical result tracks the analytic Type-1.5 rate when there is
  no gap.  The actual evolutionary sequence of the binary is
  highlighted with a thick yellow line.  }
\end{figure*}
\begin{figure*}
\centering
\mbox{\includegraphics{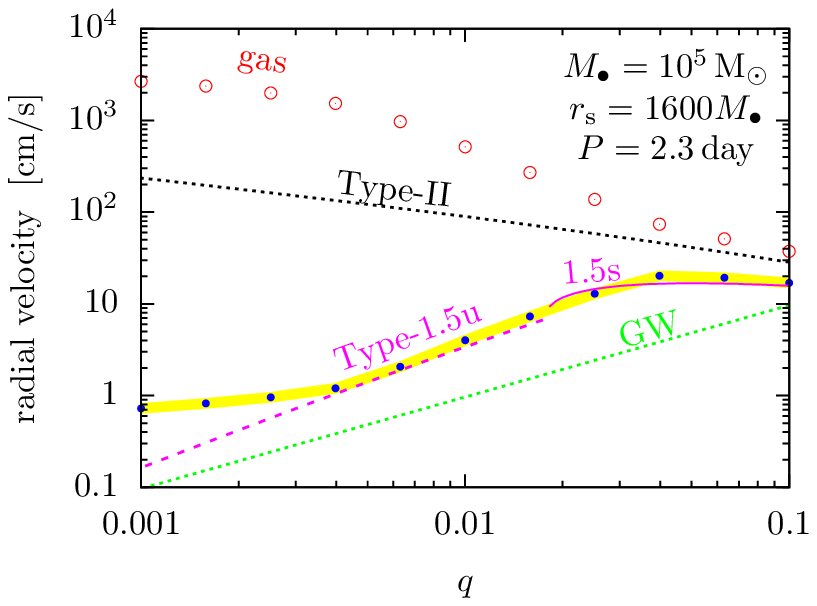}}\qquad
\mbox{\includegraphics{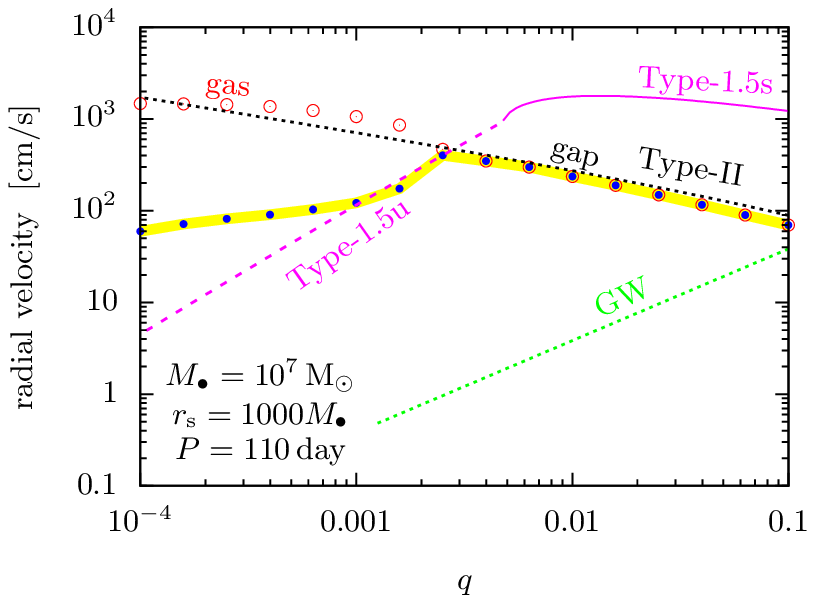}}
\caption{\label{f:migrationrate-q}
Same as Fig.~\ref{f:migrationrate-r} but varying $q$ at fixed orbital radii.
}
\end{figure*}

Next let us consider the limit that the mass ratio is sufficiently
large (i.e. more comparable masses), the tidal torque clears a gap
such that the nearby gas accretion velocity matches the migration rate
of the secondary.  If the mass of the secondary is larger than the
local disk mass then gas builds up outside the gap, reducing the
accretion velocity and increasing the migration rate until the two
match.  The resulting \textit{secondary dominated Type-II migration}
is much slower than the accretion velocity of an unperturbed
disk.

Substituting {the steady-state viscous torque}, $T_{\nu}^{\rm mg}$ from Paper~I
in Eq.~(\ref{e:vSCOr}) we get
\begin{align}
v_{\SCO r, {\rm II}} &=
-\frac{3^{3/4} \alpha^{1/2} \kappa^{1/8}}{2 \pi^{1/4}(\mu m_p/k)^{1/2}\sigma^{1/8}}
\frac{\dot{M}^{5/8}}{m_{\SCO}^{3/8} \lambda^{-11/16} \Omega_{\SCO}^{1/4} r_{\SCO}^{1/4}}\nonumber\\&=
-550\frac{\rm cm}{\rm s} \alpha_{-1}^{1/2} \dot{m}_{-1}^{5/8} M_7^{1/4} q_{-3}^{-3/8} \lambda^{-11/16}
r_{\SCO 2}^{1/8},
\label{e:v-TypeII}
\end{align}
assuming GW losses are negligible compared to the gas driven
migration.  Here $1 \lesssim \lambda\lesssim 3$ sets the relative
distance to the nearby gas outside the gap responsible for the tidal
torque (see Eq.~\ref{e:vr0}).  In Paper~I we show that depending on
whether or not the torque cutoff operates
\begin{equation}\label{e:lambdadef}
 \lambda =
\left\{
\begin{array}{l}
 1+0.13\, \alpha_{-1}^{-4/37} \dot{m}_{-1}^{4/37} M_7^{-3/37}f_{-2}^{9/37} q_{-3}^{22/37} r_{\SCO 2}^{-33/74},
\\
1+0.047\, \alpha_{-1}^{-1/3} \dot{m}_{-1}^{-1/12} M_7^{1/6} f_{-2}^{1/3} q_{-3}^{7/12} r_{\SCO 2}^{1/12}\,.
\end{array}
\right.
\end{equation}

The small blue dots in Figs.~\ref{f:migrationrate-r} and
\ref{f:migrationrate-q} show the migration rate of the secondary
$v_{\SCO r}$ and the red circles show local gas velocity
$v_{r}/\lambda$ in the numerical solution of \S~\ref{s:numerical} for
different $q$ and $r_{\SCO}$.  The dotted black curves marked as
Type-II in Figs.~\ref{f:migrationrate-r} and \ref{f:migrationrate-q}
show that the analytic formula is in excellent agreement with the
numerical steady-state solution when a gap forms.
Figure~\ref{f:migrationrate-q} shows that the migration rate is
reduced relative to the unperturbed disk gas velocity (red curves at
$q\rightarrow 0$) at the given radius.

Our result for steady-state Type-II migration, Eq.~(\ref{e:v-TypeII}),
is consistent with the secondary dominated Type-II migration rate of
\citet{1995MNRAS.277..758S}, suppressed by a factor
$\lambda^{-11/16}\lesssim 2$.  Note that other than $\dot{M}$ and this
weak $\lambda$ dependence, the Type-II migration rate is insensitive
to the details of the disk physics and the disk-satellite interaction
whenever a gap exists.  (In particular to the $f$ coefficient in the
tidal torque Eq.~(\ref{e:Lambda}), the pressure gradient, the thermal
state of the disk, and the torque cutoff.)  The Type-II migration rate
is set by the accretion rate, which is set by the boundary condition,
which we assume is a fixed fraction of the Eddington value of the
primary.

Further modifications are possible if the disk is not in steady-state.
The self-similar solution of \citet{1999MNRAS.307...79I}
{(see also \citealt{2012arXiv1205.5017R}) }
leads to a
slower migration rate (by another factor $\sim 2$ relative to
Eq.~\ref{e:v-TypeII}).  There, the steady-state accretion rate is
assumed to hold in the far zone outside the gap and equal to the rate
$\dot{M}_{\rm Edd}$ corresponding to the Eddington limit near the
primary, but the accretion rate is reduced in the middle zone relative
to the steady-state solution, as a fraction of $\dot{M}$ is
continuously used to gradually build-up the gas mass outside the gap,
which never fully reaches the steady-state level. {While these models
may be more realistic,} however, there is
no strong reason to believe that the far zone accretion rate is
related to the Eddington limit of the primary, in a scenario with a
gap where the gas is not present in the vicinity of the primary.  If
one were to assume that $\dot{M}$ is set by the inner boundary
condition in the middle zone to a given fraction of $\dot{M}_{\rm
  Edd}$, (implying that this may require a correspondingly somewhat
larger $\dot{M}$ in the far-zone), this would have led to the original,
{quasi-steady-state} \citet{1995MNRAS.277..758S} migration rate.

Another possibility was considered in \citet{2009MNRAS.398.1392L}, in
which the total amount of gas mass is limited to be less than the
secondary mass, so that the steady-state with $\dot{M}_{\rm Edd}$
cannot be reached.  They argued that the migration is much slower in
this case, and the disk cannot deliver the secondary efficiently to
separations where GW emission is sufficient to lead to merger in a
Hubble time (final parsec problem, see
\S~\ref{s:migration:GWinspiral}).

\subsection{Type-1.5 migration}\label{s:migration:type-1.5}
{
In the previous description,
Type-I migration requires azimuthally linear perturbations, which is applicable
if $r_{\rm H} \lesssim H$, implying that $q \lesssim 3\, (H/r)^3$. By contrast
Type-II migration assumes a truncated disk with negligible overflow
which is valid if the gap does not close as discussed in \S~\ref{s:gap}.
Let us now consider the intermediate Type-1.5 case,
when the secondary mass is large enough that its tidal torques cause a significant gas build--up
exterior to the secondary orbit but without completely truncating the disk.

The transition between Type-I and II migration as a function of secondary mass
was previously investigated analytically by \citet{1984Icar...60...29H} and \citet{1989ApJ...347..490W}
(see also \citealt{1997Icar..126..261W}), numerically by \citet{2003MNRAS.341..213B},
and in both ways by \citet{2007MNRAS.377.1324C},
assuming a constant sound speed and viscosity, in cases where the gas pile-up
outside the secondary is small.
Here we discuss Type-1.5 migration assuming a quasi-steady-state derived in Paper~I
in the limit of a large pile-up, self-consistently accounting for variations caused by tidal heating,
which changes the temperature, scale height, viscosity in the disk.

The migration rate is generally proportional to the dimensionless angular momentum flux
in the gas driven regime (Eq.~\ref{e:vSCOr2}).
This is sensitive to the nearby gas density and the characteristic radius
at} {which the torque is exerted, $v_{\SCO r}\propto k\propto \Sigma\Delta^{-3}$,
where $\Delta\sim \max(r_{\rm H},H)$.
If a wide gap forms, $\Delta$ is set such that $v_{\SCO r}$ matches the gas inflow rate
$v_{\SCO r}=v_{r}\propto \dot{M}/\Sigma \propto \nu$ (Type-II).
However, in the overflowing case, the migration rate becomes slower than Type-II as
the pile-up is limited by gap overflow, decreasing $\Sigma$ and $H$ compared to the
case with a gap. Relative to an unperturbed disk, although $\Sigma$ is decreased close to the secondary,
it is strongly increased due to pile-up outside of $\Delta$.
However, more importantly, the pile-up also increases the viscous stress and $H$,
which affects the migration rate greatly as $v_{\SCO r}\propto H^{-3}$ if $H>r_{\rm H}$.
Accounting for the change in $H$ is essential to correctly estimate the migration rate.
The competition of these effects determines the Type-1.5 migration
rate in the overflowing regime.

In the self-consistent steady-state overflowing model the migration rate can be obtained by
substituting $k_{\SCO}$ from Eqs.~(\ref{e:kmos}--\ref{e:kmou}) in Eq.~(\ref{e:vSCOr2}).
In the torque--saturated state,} which is most relevant for a
relatively massive secondary, the migration speed is
\begin{align}\label{e:v-Type1.5s}
 v_{\SCO r, 1.5\rm s} &=
23\frac{\rm cm}{\rm s}\, \alpha_{-1}^{-2/11} f_{-2}^{5/22} q_{-3}^{-6/11} M_7^{23/22}
r_{\SCO 2}^{83/44} \\ \nonumber
&\quad\times|\W(-a)|^{13/11}\,.
\end{align}
In the opposite, {torque--unsaturated} regime for a relatively low-mass secondary,
\begin{align}\label{e:v-Type1.5u}
 v_{\SCO r, 1.5\rm u} &=
 31\frac{\rm cm}{\rm s}\, \alpha_{-1}^{-2} M_7^{3/2} f_{-2}^{5/2} q_{-3}^{3/2}
 r_{s2}^{3/4}\left[1+\frac{\delta r_{\rm i}}{r_{\SCO}}\right]^{-35/6}\,.
\end{align}
Here GW emission emission is assumed to be negligible, appropriate in
the zone to the right of the dotted line in the $r_{\SCO}$--$q$ plane
in Fig.~\ref{f:gap} (i.e. large $r_{\SCO}$ and/or small $q$).  These
formulae apply only when the tidal torques generate a sufficiently
strong barrier that the surface density of the exterior disk is
greatly modified but a gap does not open. This requires $r_{\SCO}$ and
$q$ to fall in the zone marked by ``overflow (Type-1.5)'' in
Fig.~\ref{f:gap}.

The migration rate is not always Type-1.5 for other choices of $q$ and
$r_{\SCO}$ as shown in Figs.~\ref{f:migrationrate-r} and
\ref{f:migrationrate-q}.  However, the migration rate is approximated
to within $20\%$ by Type-1.5,
Eqs.~(\ref{e:v-Type1.5s}--\ref{e:v-Type1.5u}), for $q\gtrsim 10^{-3}$
and $r_{\SCO}\gg r_{\ISCO}$ whenever a gap is not present and GW
emission is negligible, {provided that the disk is in a quasi-steady
state and the torque model is given by Eq.~(\ref{e:Lambda}).}
These formulae capture the $(q,r_{\SCO})$
parameter dependence remarkably well in this range, where the Type-I
and II rates are inapplicable.
{Note, however, that since our results, Eqs.~(\ref{e:v-Type1.5s}--\ref{e:v-Type1.5u}),
correspond to the case of a large pile-up,
this is insufficient to investigate very small mass ratios where
the transition between the Type-I and 1.5 regimes takes place. }

Figures~\ref{f:migrationrate-r} and \ref{f:migrationrate-q} show that
Type-1.5 migration is generally slower than the other velocity scales
in the problem.  Increasing the secondary mass at a fixed orbital
radius generally increases the amount of gas mass accumulated outside
the secondary, which causes the gas velocity to be reduced for a fixed
accretion rate.  However, since there is no gap in the disk, gas still
flows in faster than the migration rate of the object.  Type-1.5
migration is slower than the secondary-dominated Type-II migration
rate, since the gas pileup in these overflowing solutions just outside
the orbit is less pronounced compared to the case with a gap.
Finally, it is also slower than the Type-I rate if the tidal torques
decrease the gas density close to the secondary or if the scaleheight is
increased.

\subsection{GW driven inspiral}\label{s:migration:GWinspiral}

As the separation decreases, the rate of energy and angular momentum
loss associated with GW emission increases rapidly. The GW-inspiral
speed (from the standard quadrupole formula for GW emission, \citealt{1964PhRv..136.1224P}) is
\begin{equation}\label{e:vGW}
 v_{\SCO r}^{\GW}=-\frac{64}{5} q \R_{\SCO}^{-3}\,.
\end{equation}
This is shown by a green dotted line in Figs.~\ref{f:migrationrate-r}
and \ref{f:migrationrate-q}.  Unlike Type-I, Type-1.5, or Type-II
migration, the GW inspiral {velocity} increases toward smaller
separations. The GW inspiral timescale is larger than the Hubble time
at large separations, where the evolution is driven by the gaseous
disk. Gas may be responsible for delivering binaries to the
separations corresponding to the GW--driven regime (but see discussion
in \S~\ref{s:migration:type-II}).  Gas-driven migration may transition
to the GW-driven regime either in the presence of a gap for comparable
mass binaries for $M\gtrsim 10^7\Msun$, or in the overflowing state
for smaller mass ratios and/or total masses (see Fig.~\ref{f:gap}).

\subsection{Summary of migration rates}

We can now summarize and interpret the trends seen for different mass
ratios and radii in Figs.~\ref{f:migrationrate-r} and
\ref{f:migrationrate-q}.  At sufficiently large radii, a gap opens for
comparable mass ratio binaries.  As gas accumulates outside the gap,
the radial gas velocity is slowed down to match the migration rate of
the secondary, according to Type-II migration.  For smaller secondary
masses in this regime, the amount of accumulated gas is less and the
corresponding gas accretion velocity becomes higher and the migration
rate faster.  At sufficiently small secondary mass and binary
separation, the tidal torque cannot keep the gas outside the Hill
sphere and the gap closes.  The gas accretion velocity in this regime
is still far slower than the unperturbed value, and the pressure and
gas density are significantly enhanced outside the orbit.  The
migration here is well approximated by the Type 1.5 rate {given by Eqs.~(\ref{e:v-Type1.5s}--\ref{e:v-Type1.5u})}. The
secondary tends to decrease the surface density and pressure within
the Hill sphere relative to the unperturbed value.  The migration rate
approaches the Type-I rate from below for very small $q$.  Along the
inward migration of the binary, the amount of local gas mass torquing
the binary decreases.  This implies that the migration rate decreases
with decreasing radius for all three cases (Type-I, 1.5, and II).  GW
emission dominates over gas driven migration in the final stages
before merger.\footnote{A separate class of migration for low-mass
  planets, which we did not consider here is Type-III migration
  associated with torques exerted by the gas on horseshoe orbits
  around the secondary \citep{2003ApJ...588..494M}.  This is
  significant if the gas mass is considerable relative to the
  secondary mass. Although the secondary mass dominates over the local
  disk mass for BH binaries embedded in accretion disks of interest
  here, future studies should investigate the effects of the gas
  crossing the secondary orbit (see the recent study by
  \citealt{2012arXiv1202.6063R} for the importance of torques due to
  gas streams in the central cavity around a SMBH binary).
}

\section{Observational implications}\label{s:observations}
\subsection{Bolometric luminosity}
\begin{figure*}
\centering
\mbox{\includegraphics{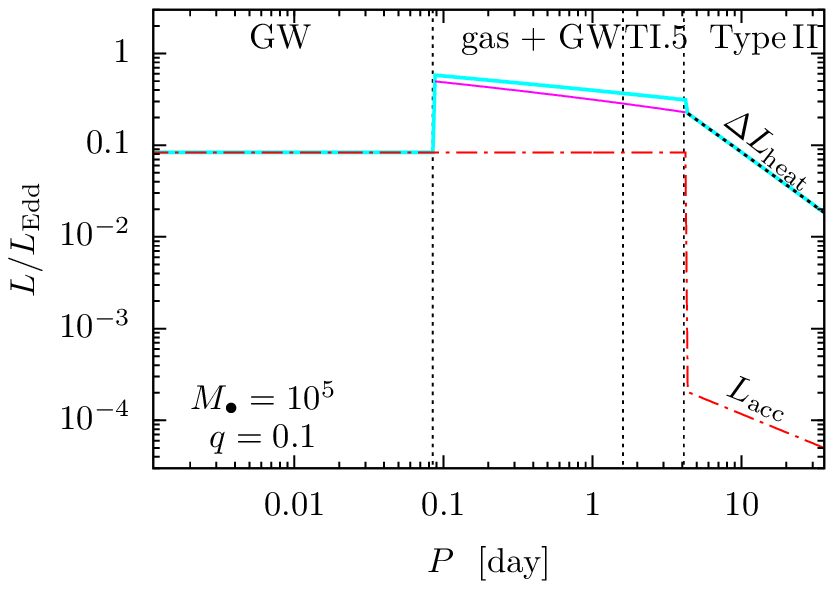}}\qquad
\mbox{\includegraphics{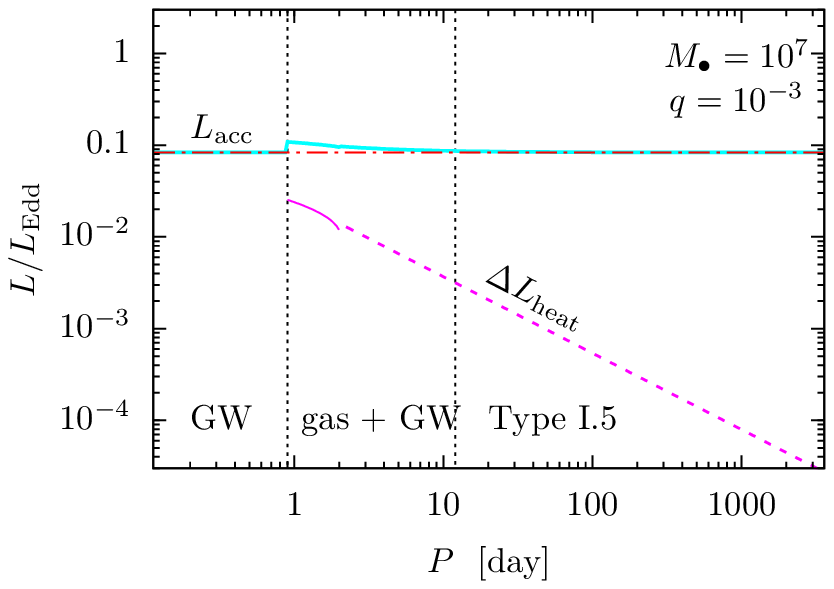}}
\caption{\label{f:lightcurve}
  The bolometric luminosity corresponding to accretion and tidal
  heating as a function of the orbital period of the secondary for
  $(M_{\SMBH},q)=(10^5\Msun,0.1)$ and $(10^7\Msun,10^{-3})$ in the
  various evolutionary stages.  The line styles for $\Delta L_{\rm
    heat}$ are the same as in Fig.~\ref{f:migrationrate-r}: black
  dotted for Type-II migration, solid and dashed magenta for saturated
  and unsaturated Type-1.5 migration.  The cyan line shows the total
  bolometric luminosity $L_{\rm acc}+\Delta L_{\rm heat}$.  Note that
  our calculations for $\Delta L_{\rm heat}$ neglects GW emission, it
  breaks down in the GW-inspiral regime.  $L_{\rm acc}$ is quenched
  when a gap opens.  }
\end{figure*}
The total luminosity of the disk is comprised of the viscous and the
tidal heating.  The energy loss corresponding to inward migration
appears as an extra source of heat in the disk, which leads to the
brightening of the disk. If the disk is radiatively efficient up to an
inner radius $r_{\min}$, then the emitted luminosity corresponds to
the energy loss\footnote{{We include factors of $\G$ and $\C$ in this section.}}
\begin{align}
 L &= \frac{1}{2} \frac{\G M_{\SMBH} \dot{M}}{r_{\min}} -  \frac{1}{2}\frac{\G M_{\SMBH} m_{\SCO}}{r_{\SCO}^2} (v_{\SCO r}-v_{\SCO r}^{\rm GW})
\\&= \frac{\dot{M} \C^2}{2\,\R_{\min}} - \frac{1}{2}\frac{\C^5}{ \G}\frac{q}{\R_{\SCO}^2 }\frac{v_{\SCO r}}{\C}
+ \frac{32}{5}\frac{\C^5}{ \G} \frac{q^{2}}{\R_{\SCO}^{5}}
\end{align}
where $\R=r/(\G \C^{-2} M_{\SMBH})$ and $v_{\SCO r}$ is the migration
rate due to gas and GW losses ($v_{\SCO r}<0$ for inward motion), and
$v_{\SCO r}^{\GW}$ is the GW inspiral rate neglecting torques from the
gas. The first term is due to accretion, the second and third terms
together are due to tidal heating, labeled $L_{\rm acc}$ and $\Delta
L_{\rm heat}$ below, respectively.

Figure \ref{f:lightcurve} shows the two components of the luminosity
$L_{\rm acc}$ and $\Delta L_{\rm heat}$ relative to $\dot{M} \C^2$ as
a function of the binary orbital time when neglecting GW
emission. Here we assume that $\dot{M} \C^2 = L_{\rm Edd} =1.4\times
10^{45} M_7\, \rm erg/sec$ where $M_7=M/10^7\Msun$. The point and line
styles are the same as in Figs.~\ref{f:migrationrate-r} and
\ref{f:migrationrate-q} in the various regimes of migration.  The
figure shows that the secondary can greatly modify the luminosity of
the disk.  The accretion luminosity is greatly decreased if a gap
forms, but the excess brightness of the outer disk, $\Delta L_{\rm
  heat}$, can overcompensate for the decrease of $L_{\rm acc}$.

The analytic results for the migration rates in \S~\ref{s:migration}
can be used to understand the features shown in
Figure~\ref{f:lightcurve}.  If a gap is opened and the secondary
exhibits Type-II migration, then the luminosity associated with
accretion scales with the orbital period $P$ as $L_{\rm acc,
  II}\propto 1/r_{\SCO} \propto P^{-2/3}$ and the secondary causes an
excess $\Delta L_{\rm heat, \rm II} \propto q^2 v_{\SCO r,
  II}/r_{\SCO}^2 \propto q^{13/8} r^{-15/8} \propto q^{13/8}
P^{-5/4}$.  In the GW-driven regime before gap decoupling, the
accretion luminosity is the same but the excess tidal heating rate is
negligible.  At smaller separations the accretion luminosity
increases, and reaches $\dot{M}\C^2/(2r_{\ISCO})$ as the gap closes.
As the binary becomes GW-driven at smaller separations the gas
build-up is decreased outside of the secondary and the excess heating
is suppressed. This transition happens in the region marked as gas+GW
driven in Figs.~\ref{f:gap}; at orbital periods
$P\lesssim (81, 14, 10 )\, \rm days$ for $q=(0.1,0.01,0.001)$,
respectively, for $M_{\SMBH}=10^7\Msun$ in particular.  Once the gap
closes, $r_{\min}$ is associated with the ISCO, and the accretion
luminosity is constant. The tidal heating in the overflowing state is
$\Delta L_{\rm heat, 1.5} = q^2 v_{\SCO r, 1.5}/r^{2}\propto q^{7/2}
r^{-5/4} \propto q^{7/2} P^{-5/6}$ in the unsaturated Type-1.5 state,
and $q^{16/11} r_{\SCO}^{-5/44} \propto q^{16/11} P^{-5/66}$ in the
saturated case.  For weakly perturbed disks where the secondary
undergoes Type-I migration, $\Delta L_{\rm heat, I}=q^2 v_{\SCO r,
  I}/r_{\SCO}^2 = q^3 r^{9/10} = q^3 P^{3/5}$ in the radiation
pressure dominated regime and $q^3 r^{-6/5} = q^3 P^{-12/15} $ in the
gas pressure dominated regime.

Figure~\ref{f:lightcurve} shows that in some cases $\Delta L_{\rm
  heat}\sim L_{\rm Edd}$. In this case, the disk near the secondary
becomes brighter than the inner accretion disk, so that the disk
appears moderately super-Eddington. While this may appear
contradictory, we demonstrate that this does not violate the local
Eddington flux limit \citep{1980ApJ...242..772A,2010ApJ...714..404T}.
As shown in \S~\ref{s:interaction}, the disk flux can be obtained
directly from the temperature and the opacity of the disk.  If the
disk is supported vertically by gas and radiation pressure, this can
be expressed in terms of the disk scaleheight. Assuming that the sound
speed satisfies $H=c_{\SCO}/\Omega$, the flux is given by $F =
(\C/\kappa) H \Omega^2 (1-\beta)$ (see Paper I). For a nearly
Keplerian thin disk, the integrated flux from the two faces is then
\begin{align}\label{e:Ltot}
 L = \frac{4\pi G M_{\SMBH} \C}{\kappa} \int_{r_{\min}}^{r_{\max}} (1-\beta)\frac{H}{r} \frac{\D r}{r}\,.
\end{align}
Here the first factor outside the integral can be identified as the
Eddington luminosity $L_{\rm Edd}$ for a source in hydrostatic
equilibrium, while the integral is a geometrical factor which is less
than unity for a thin disk truncated at the ISCO. For a stationary
accretion disk without a secondary, $H(r)$ is a constant in the most
luminous, radiation pressure dominated regime ($\beta\approx 0$), and
the integral reduces to $L=\dot{M}\C^2/(2 \R_{\min})$.  More
generally, $L=\eta \dot{M}\C^2 = \epsilon L_{\rm Edd}$. In
Fig.~\ref{f:lightcurve}, we conservatively chose $\R_{\min}=6$
corresponding to a non-spinning BH, and $\epsilon = \eta$ consistent
with AGN observations showing that $\epsilon\sim 10\%$--$25\%$ for
bright AGNs \citep{2006ApJ...648..128K,2009ApJ...700...49T}.  The
luminosity increase due the secondary is significant if $H/r$ is
larger near the secondary than near the ISCO and if the disk is
radiation pressure dominated.  Indeed, Fig.~\ref{f:disk} has shown
that $H/r$ can increase dramatically for $q>10^{-3}$.
Eq.~(\ref{e:Ltot}) shows that a thick, radiation pressure dominated
accretion disk can exceed the hydrostatic Eddington limit by a
logarithmic factor, $\ln(r_{\max}/r_{\min})$. This is due to the
velocity shear and vorticity in the disk, which are neglected in the
hydrostatic Eddington limit \citep{1980ApJ...242..772A,2004astro.ph..11185A}.

We conclude that the disk luminosity may be modified significantly by
the orbiting object. Whenever a gap forms, the disk becomes either
fainter due to the loss of accretion onto the primary or brighter due
to the accumulated gas mass outside the secondary heated by tidal
dissipation (with the latter effect dominating near the end of the
Type II migration phase in many cases).
The competition of these effects generates a unique lightcurve which
is sensitive to the type of migration as well as to GW losses.  These
conclusions assume axisymmetry; an obvious caveat is that
non-axisymmetric accretion streams could generate significant
luminosity \citep{2008ApJ...682.1134H,2011MNRAS.415.3033R,2012arXiv1202.6063R}.

\subsection{Disk spectrum}
\begin{figure*}
\centering
\mbox{\includegraphics[width=8.5cm]{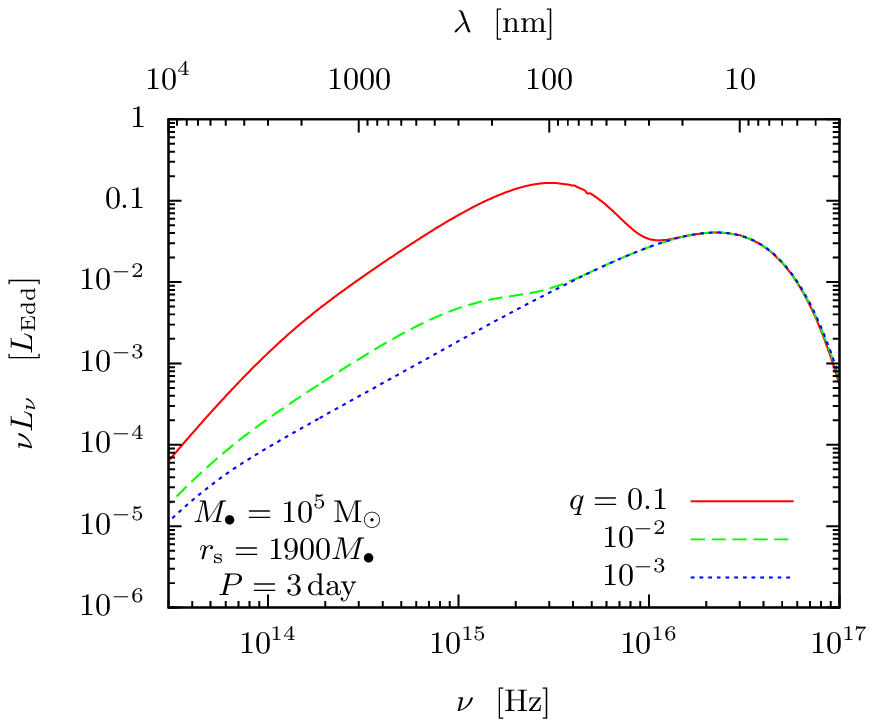}}\quad
\mbox{\includegraphics[width=8.5cm]{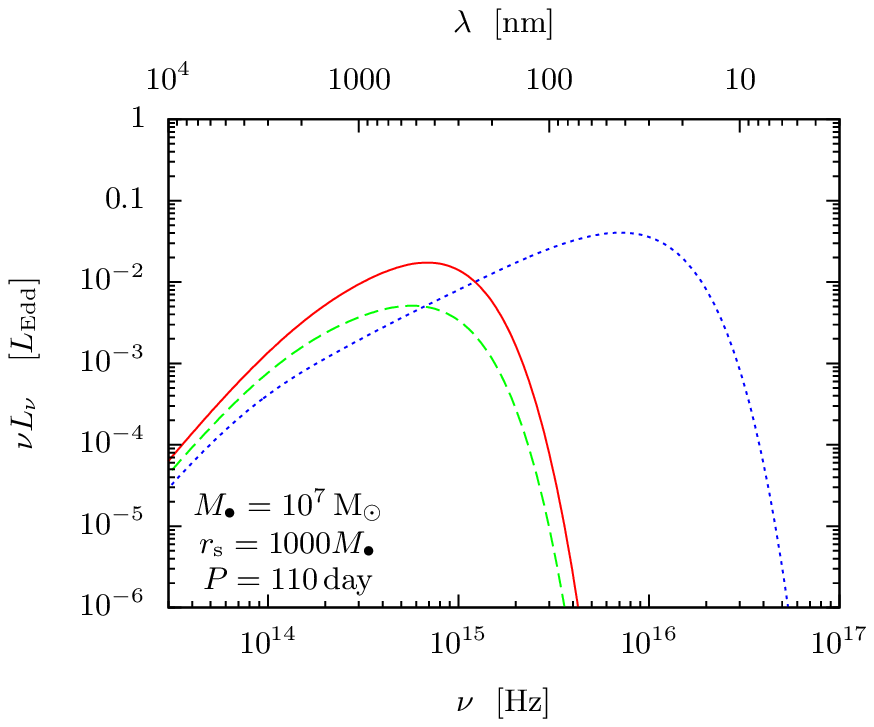}}
\caption{\label{f:disk-spec}
  Disk spectra in units of $L_{\rm Edd}\sim 10^{43}$ and
  $10^{45}{\,\rm erg/s}$ for $M_{\SMBH}=10^5$ and $10^7\,M_{\SMBH}$
  binaries with periods as marked. Different curves correspond to
  different mass ratios as labeled. The dotted curve ($q=10^{-3}$)
  represents the asymptotic spectrum of a solitary disk.  The spectrum
  on the right panel is truncated for $q=0.1$ and $0.01$ when a gap
  forms. The optical enhancement at lower frequencies is due to the
  outer disk.  }
\end{figure*}
We obtain a simple estimate of the disk spectrum by assuming multicolor black body radiation for each annulus
with the given surface temperature profile $T_{s}(r)$,
\begin{equation}
 B(\nu, T) = \frac{2 {\rm h} \nu^3 }{ \C^2} \frac{1}{e^{{\rm h} \nu/ {\rm k} T} -1}
\end{equation}
and integrating over radius
\begin{equation}
 L_{\nu} = 4\pi^2 \int_{r_{\min}}^{r_{\max}} B[\nu, T_{s}(r) ] \, r \,\D r\,.
\end{equation}
We ignore gravitational redshift and Doppler boost for simplicity
since $\R\gg 1$ (however, see discussion below).

The outer parts of the accretion disk have a much lower temperature
than the inner parts. Therefore, the brightening caused by a secondary
affects the spectrum more prominently at shorter wavelengths.  The
disk spectrum can be used to disentangle the effects of a secondary
from the properties of the inner disk.

Figure~\ref{f:disk-spec} shows the disk spectra in two representative
cases when the orbital period is $P = 3$ or $110\,{\rm days}$ for
$M_{\SMBH}=10^5$ and $10^7\Msun$, respectively. Different curves show
different mass ratios.  The spectrum is truncated at high frequencies
when a gap forms, as the hottest central regions are removed from the
disk. Relative to an accretion disk with no secondary, the disk is
much brighter in the ultraviolet, optical, and infrared bands for large secondary
masses, both in the gap forming and in the overflowing cases.  The
extra energy originates from the orbital energy of the binary which
heats the disk through the gravitational torques on top of the viscous
heating.  For other binary masses and separations assuming the binary
is in the gas driven regime, the {dimensionless angular momentum flux or } brightening factor is shown in
Fig.~\ref{f:k}. For smaller binary separations when the GW inspiral
overtakes the accretion rate, the optical and infrared enhancement
associated with the exterior gas pileup goes away.

Note that our estimates neglect accretion onto the secondary which can
generate an additional spectral peak in X-rays
\citep{2012MNRAS.420..860S,2012MNRAS.420..705T}.  We also neglected
relativistic corrections to the disk spectrum, and other features such
as spectral lines and the high-frequency tail associated with the
corona. Since the orbital velocity of the secondary on a circular
orbit is $v_{\rm orb}=3\times 10^4\, {\rm km}\,{\rm s}^{-1}\,
r_{2}^{-1/2}$, the offset and broadening of spectral lines can be
quite significant for SMBH binaries
\citep{2009ApJ...700.1952H,2010ApJ...725..249S,2012AdAst2012E...3D,McKernan12}.
Thus, the actual AGN binary spectra shown in Fig.~\ref{f:disk-spec}
may also exhibit relativistically broadened and/or offset spectral
lines by $6,900$ and $9,500 \, {\rm km}\,{\rm s}^{-1}$ in the left and
right panels, respectively. The lightcurve is expected to be
periodically modulated on the binary orbital period (3 and 110 days
respectively in the two panels; see also discussion below), with
harmonics appearing at $\sim 3$ and $0.5$ times the orbital period for
mass ratios $q\gtrsim 0.05$ (D'Orazio et al. 2012, in preparation).
These signatures together can help future observational efforts to
identify SMBH binaries in the overflowing regime.

\subsection{Transient and periodic variability statistics}
\begin{figure*}
\centering
\mbox{\includegraphics{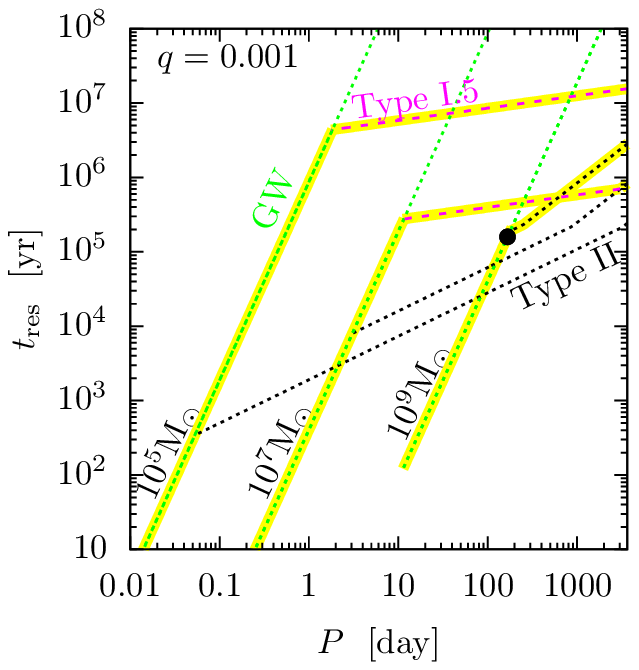}}
\mbox{\includegraphics{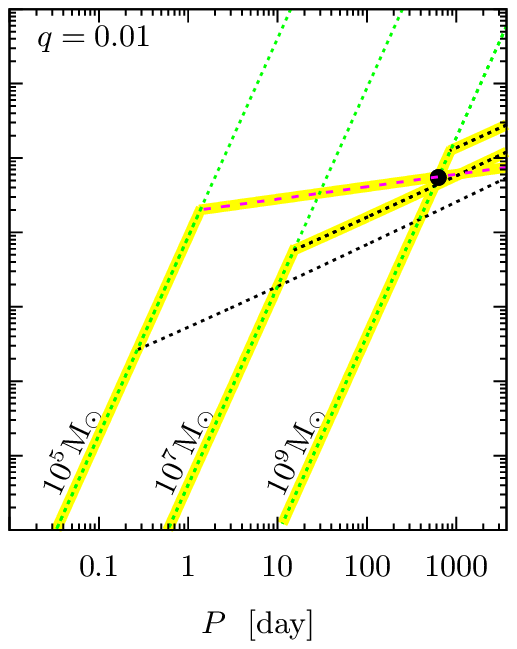}}
\mbox{\includegraphics{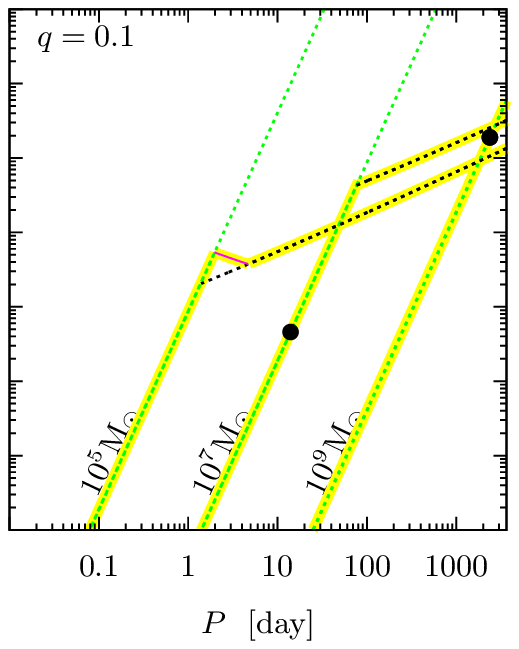}}
\caption{\label{f:tres}
  The amount of time the binary spends at various orbital periods
  during its evolution towards merger.  The three panels show
  different mass ratios, and in each panel the three sets of curves
  show different total masses as marked.  For each mass, different
  line styles correspond to different types of migration, GW inspiral
  (green), saturated and unsaturated Type-1.5 (magenta dashed and
  solid), and Type-II (black dotted).  The evolutionary tracks are
  highlighted with a thick yellow line. Big black dots mark gap
  decoupling; for much smaller periods, the tidal torques may not
  generate EM variability.  }
\end{figure*}

We have calculated the rate at which the binary evolves toward merger
and have shown that the state of the disk changes during this
process. From this one can predict the residence time $t_{\rm
  res}\equiv r/v_{\SCO r}$ the binary resides in the various states.
Assuming that the luminosity of the accretion disk is modulated
periodically on the orbital timescale {\citep{Hayasaki+07,2008ApJ...672...83M,2009MNRAS.393.1423C}}, it is possible to predict the
relative fraction of binaries that exhibit variability as a function
of orbital period \citep{2009ApJ...700.1952H}. One may expect to see
EM transients when the binary is surrounded by gas, and a different
distinct population of dimmer or softer sources decoupled from the
disk assuming that the GW inspiral dominates the evolution inside of a
gap in the disk. The predictions can be observationally tested using
large deep variability surveys, such as PanSTARRS or LSST.

Previously, \citet{2009ApJ...700.1952H} has carried out this exercise
assuming a GW inspiral at short periods and Type-II migration for long
periods.  These predictions have to be revised in two respects.
First, many of the binaries may be in the Type-1.5 regime which has a
longer residence time.
Second, we predict that tidal torques cannot sustain a gap once a
sufficient amount of gas has accumulated, which may cause
gap-refilling, before the binary can ``run away'' from the gap edge in
the GW-driven inspiral regime. Without gap-refilling or an inner disk
around the primary and secondary, periodic variability is not expected
in the GW-driven regime with a decoupled outer disk (see however
\citealt{2011PhRvD..84b4024F,2012ApJ...744...45B,2012arXiv1204.1073N,2012arXiv1207.3354F}).  Future studies
should investigate the expected variability in circumbinary or
overflowing disks, where the secondary is in the GW-driven regime.

Figure~\ref{f:tres} shows the residence time the binary spends at
various orbital periods for different masses and mass
ratios. Different line styles show the type of migration.  The
residence time increases with period very quickly ($t_{\rm res}\propto
t_{\rm orb}^{8/3}$) in the GW driven regime, but not so rapidly in the
gas driven regime. The dependence is flatter for Type-II, and even
more so for Type-1.5, where it can even decrease with period in the
torque-cutoff state (at $P\sim 1\,$day for $10^5\Msun$ in the right
panel).  This shows that the efficiency at which gas can deliver
objects to the GW driven regime is deteriorated by gas overflow. The
transitions to GW-driven inspiral occurs at longer periods for the
Type-1.5 than for Type-II migration (see
Fig.~\ref{f:migrationrate-r}).  The fraction of binaries at orbital
periods between a few days to years is larger if Type-1.5 migration
operates, relative to the Type-II rate. As a rule of thumb, we find
this to be the case if one of the black holes is less massive than
$10^5\Msun$.  Conversely, the residence time follows the Type-II rate
all the way to the GW inspiral regime if both objects are more massive
than $10^5\Msun$.

Gap refilling may be witnessed through the increase in the bolometric
luminosity and spectral X-ray hardening of the source
\citep{2005ApJ...622L..93M,2010AJ....140..642T}.  Our results show
that this may occur already before merger, particularly for
$10^5\Msun$ SMBH masses in the {\it LISA/NGO} range \citep{2005ApJ...623...23S}, generating bright
electromagnetic sources coincident with GWs.  Gap refilling may occur
more rapidly
compared to the viscous time without the pile-up, due to the enhanced
gas mass and stress outside the gap and the presence of the secondary,
implying a larger population of birthing quasars for all-sky optical
(e.g. LSST) or soft X-ray surveys.  However, since this occurs at
larger radii where the viscous time is longer, it is unclear whether
this yields a larger or smaller population of birthing quasars with
significant brightening during the mission lifetime of all-sky optical
(e.g. LSST) or soft X-ray surveys.

The significant excess flux in the gas driven regime may increase the
prospects for identifying SMBH binary sources. Future surveys for
these sources can {provide observational evidence for SMBH mergers in the
{\it NGO} mass range or IMBHs orbiting around SMBHs.}

\subsection{Gravitational wave measurements}

Gap refilling and Type-1.5 migration has implications for GW
measurements in several ways.  First, we have shown that the gap
closes much earlier than previously thought. This implies that the
{\it LISA/NGO} binaries may be embedded in gas even for nearly comparable mass
ratios.  This helps the prospects of identifying coincident EM
counterparts to {\it LISA} sources.  Second, the effects of gas may be
identified directly from the GW signal itself.  The GWs emitted by
SMBH binaries is in the frequency bands detectable by pulsar timing
arrays (PTAs) if the orbital period is between a few days to a few
years, and in the planned {\it LISA} range if it is between a few seconds
to a $\sim$half day. Figure~\ref{f:tres} shows that the corresponding
mass range is below $10^6\Msun$ for {\it LISA} and above $10^7\Msun$ for
PTAs. Here we describe how the accretion disk affects the GW signal
for {\it LISA} and PTAs, and discuss the implications of gas overflow and
Type-1.5 migration.

\subsubsection{PTAs}

The nHz GW background, measurable by PTAs, is generated by SMBH
binaries {\citep{2008MNRAS.390..192S,2009MNRAS.394.2255S}}, which emit a stationary signal with angular frequency
$2\Omega$ for circular sources and many distinct upper harmonics for
eccentric sources. The background is affected by the disk in two ways:
(i) by changing the relative fraction of binaries at particular
orbital radii according to the residence time
\citep{2011MNRAS.411.1467K}, and (ii) by changing the eccentricity
distribution of sources if a gap is opened
\citep{1992PASP..104..769A,2005ApJ...634..921A,2009MNRAS.393.1423C,2011MNRAS.415.3033R,2012arXiv1202.6063R}.

\citet{2011MNRAS.411.1467K} calculated the GW background for a
population of binaries that undergo Type-II migration in steady-state
$\alpha$ or $\beta$ disks, or the self-similar migration according to
the quasi-stationary model in \citet{1999MNRAS.307...79I}. They found
that the unresolved GW background is typically not reduced
significantly in the PTA frequency band if the sources undergo
secondary dominated Type-II migration.\footnote{The background could
  be reduced only if Type-II migration occurs on the viscous timescale
  with no accumulation of gas.}  This is due to the fact that the
background is dominated by comparable mass binaries for which (1) the
slowdown of Type-II migration in the secondary dominated regime is
more significant and (2) many of these sources transition to a GW
dominated evolution at large orbital periods outside the PTA range.
Regarding the individually resolvable GW sources with PTAs, they are
typically the most massive $q\sim 1$, $\sim 10^9\Msun$ binaries, which
are not affected by gas in any case at these frequencies.
Figure~\ref{f:tres} shows that Type--1.5 migration in $\beta$--disks
is not relevant for PTA measurements since it does not affect the
binaries with masses larger than $10^6\Msun$. Since gaps are expected
not to close for these masses for $\beta$--disks, these sources are
expected to exhibit GW spectra characteristic of eccentric
sources. Further studies should examine whether gap overflow may occur
for PTA sources embedded in $\alpha$ disks, and the implications for
eccentricity.

\subsubsection{LISA/NGO}

As mentioned above, the frequency range of {\it LISA/NGO} corresponds to binary
orbital periods shorter than a day, the sensitivity is best for $P
\sim 10\,\min$. In this regime, the binaries are typically already in
the GW inspiral regime. Nevertheless, the effects of gas are imprinted
on the frequency and phase evolution of the signal as the orbital
period shrinks during the measurement. In \citet{2011PhRvD..84b4032K}
and \citet{2011PhRvL.107q1103Y}, we have shown that the corresponding
GW phase perturbation is very significant ($\sim1000\,$rad/yr)
relative to the {\it LISA} measurement accuracy ($\sim1\,$rad/yr) if the
gaseous torque corresponds to secondary dominated Type-II migration
(particularly for $M_{\SMBH}=10^6\Msun$), and less pronounced
($\sim10\,$rad/yr) but still significant if it corresponds to Type-I
migration.  Here we have not examined the gaseous torques in the
GW-driven regime.  In that case, since the viscous inflow is slower
than the GW-inspiral, the gas does not bank-up outside of the
secondary.\footnote{Gas may bank up interior to the orbit in the GW
  driven regime \citep{2010MNRAS.407.2007C}, see however
  \citet{2012MNRAS.tmpL.436B} and \S~\ref{s:literature} above.}
Therefore, we conclude that gas effects are expected to be reduced
compared to the Type-II case in \citet{2011PhRvL.107q1103Y}, but
future studies should examine whether the corresponding phase shift is
still measurable with the {\it LISA} accuracy.

The GW spectrum may also be affected by the binary eccentricity,
reminiscent of gas \citep{2002ApJ...567L...9A}. We have shown that the
disk overflows for typical {\it LISA} binaries. Further studies should
investigate whether the eccentricity is excited in the overflowing
state, and whether this remains significant in the GW inspiral regime
when the binary enters the {\it LISA} frequency range.

\section{Conclusions}\label{s:conclusions}

We have examined the self-consistent steady-state structure of
accretion disks around SMBH binaries, conditions of gap closing, and
migration rates of the binaries. Our main conclusions can be
summarized as follows.

\bigskip
\begin{enumerate}
\item \textit{Gap closing---} The secondary represents a strong tidal
  barrier in the accretion disk, which causes a significant
  accumulation of gas in an extended range outside of the secondary.
  Still, the tidal torque cannot counteract the enhanced viscous
  torque in steady-state for a large range of binary masses,
  particularly for $\lesssim 10^5\Msun$, and the disk does not form a
  gap.  The steady-state exhibits continuous overflow across the
  orbit.
\item \textit{Phase diagram---} The disk is described by separate
  formulas in the two main regions in the binary parameter space when
  GW emission is negligible: A gap is present for sufficiently large
  separations and masses (Fig.~\ref{f:k}) but typically refills before
  GW-driven decoupling.  We identify further sub-regions in this
  ``phase space'' depending on whether the tidal torques are
  saturated.
\item \textit{Type 1.5 migration---} The gas-driven migration rate is
  Type-I for weakly perturbed disks, Type-II where the gap can remain
  open, and Type-1.5 in the intermediate strongly perturbed inflowing
  state. We have derived analytic formulas for the Type-1.5 rates and
  have found these to be slow compared to both Type I and II rates.
\item \textit{optical and infrared excess---} The disk flux is
  increased by a factor of up to $100$--$500$ due to gas pileup
  outside the orbit of the secondary for binaries with
  $M_{\SMBH}=10^5$--$10^7\Msun$ in the Type-1.5 and Type-II migration
  regimes, enhancing the ultraviolet, optical, and near infrared brightness of the
  disk. The enhancement is more pronounced for lower $M_{\SMBH}$.  The
  disk can become moderately super-Eddington at these frequencies.
  The excess brightness is mitigated by gas overflow and further suppressed
  for binaries for which GW-emission is significant.
  This can be used to indirectly indicate the presence of GWs.
\item \textit{Periodically variable AGN---} The orbital period is
  between 1 day and 10 years for binaries in the Type-1.5 migration
  regime. Periodic variability surveys of AGNs can discover these
  sources and test the predicted spectral signatures.  The statistics
  of the relative abundance of many such sources with different
  periods can be used to observationally test the migration and GW
  inspiral rate, and to estimate the expected {\it LISA} merger rate.
\item \textit{EM counterparts for {\it LISA}---} The masses most affected
  by gap overflow and Type--1.5 migration is in the {\it LISA} range (near
  $\sim 10^5\Msun$). The gap closes even for comparable mass ratios
  before GW emission dominates, implying that normal AGN-like activity
  may be coincident with {\it LISA} sources {\citep{2006ApJ...637...27K}}.
\item \textit{PTA sources---} The gap can remain open for the massive
  $10^8$--$10^9\Msun$ binaries, implying that the GW background is not
  affected by Type-1.5 migration for PTAs. The presence of gas can be
  inferred either from EM surveys or indirectly from the GW spectrum,
  since the gas increases the binary eccentricity if a gap is open,
  which adds orbital harmonics to the spectrum. {The eccentricity is
  expected to be much smaller for systems devoid of gas
\citep{2010ApJ...719..851S,2011ApJ...732L..26P,2012ApJ...754...42M}.}
\item \textit{Star formation and final parsec problem---} Previously,
  \citet{2009MNRAS.398.1392L} argued that the migration rate is
  greatly reduced by star formation feedback
  if the disk is unstable to gravitational fragmentation
  and the disk mass is smaller than the secondary mass.
  However, for the low-mass, but continuously replenished disks considered here
  we find that the disk is stabilized against gravitational
  fragmentation.  This is because viscous heating increases the sound
  speed over a large range of radii outside the secondary's orbit, and
  this effect is more important than the increase of surface density
  in the same region.
\end{enumerate}

We note that all of these findings correspond to a disk model in which
the effective viscosity is proportional to the gas pressure in the
disk (so-called $\beta$-disks). Future studies should investigate
alternative models in which the viscosity is proportional to the total
gas+radiation pressure. We also assumed steady-state models, where the
accretion rate is constant, and set by the Eddington limit of the
primary.  While we expect these assumptions to be justified in the
overflowing regime, where the gas inflow rate is much higher than the
migration rate of the secondary, the steady-state Eddington accretion
assumption is suspect for circumbinary disks with gaps
\citep{1999MNRAS.307...79I,2009MNRAS.398.1392L,2012arXiv1205.5017R}.  Future studies
should address comparable mass binaries where the disk may be
significantly non-axisymmetric
\citep{2008ApJ...672...83M,2009MNRAS.393.1423C} and where the
accretion of the secondary is non-negligible
\citep{1999ApJ...526.1001L}, and examine whether binary eccentricity
is excited in the overflowing steady-state
\citep{1992PASP..104..769A,2005ApJ...634..921A,2009PASJ...61...65H,2011MNRAS.415.3033R,2012arXiv1202.6063R}.

\section*{Acknowledgments}

We thank Re'em Sari, Taka Tanaka, {Roman Rafikov, and Alberto Sesana} for useful discussions.
BK acknowledges support from NASA through Einstein Postdoctoral
Fellowship Award Number PF9-00063 issued by the Chandra X-ray
Observatory Center, which is operated by the Smithsonian Astrophysical
Observatory for and on behalf of the National Aeronautics Space
Administration under contract NAS8-03060.  This work was supported in
part by NSF grant AST-0907890 and NASA grants NNX08AL43G and
NNA09DB30A (to AL) and NASA grant NNX11AE05G (to ZH).

\bibliography{paper-2}

\end{document}